\documentclass[a4paper,fleqn]{cas-dc}
\usepackage[utf8]{inputenc}
\DeclareUnicodeCharacter{2212}{\textminus}

\usepackage[numbers,sort&compress]{natbib}
\usepackage{booktabs}
\usepackage{graphicx}
\usepackage{tabularx}
\usepackage{appendix}

\usepackage{array}
\usepackage{float}
\usepackage{placeins}
\usepackage{flushend}
\usepackage{subcaption}
\newcolumntype{Y}{>{\centering\arraybackslash}X}
\newcommand{\Ang}{\ensuremath{\mathring{\mathrm{A}}}}
\newcommand{\VperA}{\ensuremath{\,\mathrm{V}/\Ang}}

\def\tsc#1{\csdef{#1}{\textsc{\lowercase{#1}}\xspace}}
\tsc{WGM}
\tsc{QE}
\tsc{EP}
\tsc{PMS}
\tsc{BEC}
\tsc{DE}

\begin{document}
\let\WriteBookmarks\relax
\def\floatpagepagefraction{1}
\def\textpagefraction{.001}

\shorttitle{Enhanced Valley Polarization}

\shortauthors{Islam, Chowdhury, Nure-Alam-Dipu and Zubair}

\title [mode = title]{Strain- and Electric-Field-Tunable Valley Polarization in Mo\textsubscript{0.75}V\textsubscript{0.25}Te\textsubscript{2}(Mo\textsubscript{3}VTe\textsubscript{8}) for Valleytronic Application}                      


%
\author[1]{Md. Mostaqul Islam}[orcid=0009-0000-7767-0545]
\fnmark[1]
\credit{Formal Analysis, Visualization, Investigation, Writing -- original draft}

\author[1]{Vivek Chowdhury}[orcid=0009-0006-6180-8398]
\fnmark[1]
\credit{Formal Analysis, Visualization, Investigation, Writing -- original draft}

\author[1]{Md. Nure-Alam-Dipu}
\fnmark[1]
\credit{Conceptualization, Methodology, Visualization, Software}

\author[1]{Ahmed Zubair}[orcid=0000-0002-1833-2244]
\cormark[1]
\credit{Supervision, Conceptualization, Methodology, Project administration,
Resources, Funding acquisition, Writing -- review \& editing}

\affiliation[1]{
    organization={Department of Electrical and Electronic Engineering},
    addressline={Bangladesh University of Engineering and Technology (BUET)}, 
    city={Dhaka},
    postcode={1205}, 
    country={Bangladesh}
}

\fntext[1]{Md. Mostaqul Islam, Vivek Chowdhury, and Md. Nure-Alam-Dipu contributed equally to this work.}

\cortext[cor1]{Corresponding author. E-mail address: ahmedzubair@eee.buet.ac.bd (A. Zubair)}

\ExplSyntaxOn
\seq_gclear:N \g_stm_prelimsau_seq

\seq_gput_right:Nn \g_stm_prelimsau_seq
  { Md.~Mostaqul~Islam\textsuperscript{a,1} }

\seq_gput_right:Nn \g_stm_prelimsau_seq
  { \space Vivek~Chowdhury\textsuperscript{a,1} }

\seq_gput_right:Nn \g_stm_prelimsau_seq
  { \space Md.~Nure-Alam-Dipu\textsuperscript{a,1} }

\seq_gput_right:Nn \g_stm_prelimsau_seq
  { \space Ahmed~Zubair\textsuperscript{a,*} }

\ExplSyntaxOff


\begin{abstract}
Valley polarization in 2D TMDs is promising for low-power valleytronic and spin-valley information processing, but time-reversal symmetry in pristine nonmagnetic TMDs keeps the K+ and K− valleys degenerate, limiting device applications. In this work, we investigated the structural stability, electronic properties, and tunable valley polarization of V-alloyed MoTe$_2$ monolayer, Mo$_{0.75}$V$_{0.25}$Te$_2$, using first-principles density functional theory (DFT) calculations. Substitutional alloying of MoTe$_2$ with V introduced magnetic exchange interaction, which, together with spin-orbit coupling (SOC), lifted the valley degeneracy at the unequal valleys. The alloyed structure was found to be energetically and dynamically stable due to the absence of imaginary phonon modes. In pristine MoTe$_2$, SOC produced spin splittings of 34.0 meV and 218.9 meV in the conduction bands and valence bands, respectively, but no valley polarization was observed. In contrast, Mo$_{0.75}$V$_{0.25}$Te$_2$ exhibited spontaneous valley polarization of 37.3 meV in the conduction band and 78.2 meV in the valence band. The valley polarization was further enhanced by external electric fields and biaxial strain. A transverse electric field along the crystal $c$ axis produced the maximum valley splitting of 132.8 meV in the valence band, whereas biaxial tensile strain increased the valence band valley splitting up to 160.8 meV. The maximum conduction band valley splitting reached 54.4 meV under 2\% biaxial compressive strain. These results demonstrated that V alloying, combined with electric-field and strain engineering, provides an effective strategy for achieving large and tunable valley polarization in MoTe$_2$. Thus, Mo$_{0.75}$V$_{0.25}$Te$_2$ can be considered a promising 2D platform for tunable valleytronic device applications, such as transistors and sensors.
\end{abstract}








\maketitle

\section{Introduction}

Two-dimensional (2D) transition metal dichalcogenides (TMDs) have attracted promising attention due to their exceptional electronic, optical, mechanical, magnetic, and valley-dependent properties. Monolayer TMDs, particularly those with hexagonal symmetry, exhibit strong quantum confinement, sizable spin-orbit coupling (SOC), and broken spatial inversion symmetry, which make them promising candidates for next-generation electronic~\cite{wang2024critical}, optoelectronic~\cite{baugher2014optoelectronic, ciarrocchi2022excitonic}, photonic~\cite{zhang2014electrically}, and valleytronic devices~\cite{liu2019valleytronics,ciarrocchi2022excitonic}. In contrast to conventional charge-based electronics, which suffer from increasing power dissipation and scaling limitations~\cite{dennard2003design,horowitz2005ieee}.

Valleytronics employs the valley degree of freedom associated with the inequivalent K+ and K-- points at the corners of the hexagonal Brillouin zone. Similar to charge and spin, the valley index can act as an information carrier, enabling the realization of valley filters and valves~\cite{zang2017valleytronics, PhysRevB.105.L081115}, optical switches~\cite{mak2018light, PhysRevApplied.19.034056}, magnetic switches~\cite{pal2023quantum}, and nonvolatile memory devices~\cite{zang2017valleytronics,pal2023quantum}. The large momentum-space separation between the K+ and the K-- valleys suppresses intervalley scattering from long-wavelength phonons and smooth disorder, making valley-polarized states attractive for robust information storage and transport~\cite{schaibley2016valleytronics,vitale2018valleytronics}. In pristine nonmagnetic TMDs, however, time-reversal symmetry (TRS) enforces energy degeneracy between the K+ and the K-- valleys, preventing spontaneous valley polarization even when SOC induces spin splitting within distinct valleys.

Breaking the valley degeneracy is therefore essential for practical valleytronic applications. Several approaches were proposed to lift the degeneracy between the K+ and the K-- valleys, including external magnetic fields~\cite{smolenski2016tuning,zhao2017enhanced}, magnetic proximity effects~\cite{hu2020manipulation}, optical excitation~\cite{ye2017optical}, optical Stark effects~\cite{srivastava2015valley}, magnetic doping~\cite{JanusMOSSemagneticdoping}, strain engineering~\cite{vse2vn, samrat2026}, twisting~\cite{ge2022enhanced}, and external electric fields~\cite{wang2021valley}. Among these, electric field-induced valley manipulation is particularly attractive for device operation because it can be externally controlled and integrated into gate-tunable architectures~\cite{wang2021valley}. Nevertheless, many external field-based strategies suffer from important limitations. Magnetic field-induced valley splitting is generally weak; magnetic substrates may alter the intrinsic electronic structure of the active 2D layer, and optical pumping requires complex excitation conditions with short carrier lifetimes~\cite{mao2019biaxial,mak2014valley}. So, identifying 2D systems that possess intrinsic or easily tunable valley polarization remains a major challenge.

Ferrovalley materials provide an appealing pathway to overcome this limitation because they combine broken TRS, broken spatial inversion symmetry, and strong SOC in a single 2D platform. In such materials, intrinsic exchange interaction produces spin polarization, while SOC couples the spin and valley degrees of freedom, resulting in spontaneous valley splitting without the continuous application of an external magnetic field~\cite{tong2016concepts}. Several 2D ferrovalley candidates, including V-based chalcogenides, rare-earth halides, and other magnetic monolayers, were theoretically reported~\cite{song2018spontaneous,zhao2019single,ding2021prediction,cheng2021two,wu2023realizing,zhao2022intrinsic,samiul2026165094}. However, many of these systems still exhibit relatively small valley splitting, limited tunability, or structural and experimental constraints, which restrict their practical implementation in room-temperature valleytronic and spintronic devices.

Magnetic alloying offers another promising strategy for engineering spin and valley properties in otherwise nonmagnetic TMDs. Pristine group-VI TMDs such as MoS$_2$, MoSe$_2$, WS$_2$, WSe$_2$, and MoTe$_2$ are generally nonmagnetic because they do not contain unpaired magnetic moments or unsaturated bonds. Introducing transition-metal atoms with partially filled $d$ orbitals can induce magnetism, exchange splitting, and spin-polarized electronic states while preserving the atomically thin nature of the host material. Previous first-principles studies showed that alloying WSe$_2$ with VSe$_2$ induced long-range ferromagnetic order and produced spin-filtering behavior in W$_{1-x}$V$_x$Se$_2$ monolayer and VSe$_2$/VN heterostructure~\cite{hoque2022first,vse2vn}. Such results indicate that V incorporation can be an effective route to introduce magnetic ordering and spin-dependent transport in 2D TMD alloys.

Among TMDs, MoTe$_2$ is particularly interesting because the heavier Te atoms enhance SOC, while the hexagonal lattice naturally hosts inequivalent K+ and K-- valleys. Although pristine monolayer MoTe$_2$ exhibits SOC-induced spin splitting, the K+ and K-- valleys remain energetically degenerate due to preserved TRS. Pristine MoTe$_2$ alone is insufficient for spontaneous valley polarization. Substitutional alloying with V atoms can break TRS by introducing magnetic exchange interaction, while the strong SOC of MoTe$_2$ can couple this magnetic ordering to the valley degree of freedom. The combined action of exchange interaction and SOC is expected to lift the K+ and the K--  valley degeneracy and generate valley-polarized electronic states suitable for anomalous valley Hall transport~\cite{tong2016concepts}.

In addition to intrinsic alloy-induced valley polarization, external perturbations such as strain and electric fields provide powerful tuning knobs for controlling the valley splitting. Mechanical strain can modify bond lengths, bond angles, orbital hybridization, and crystal-field environments, thereby changing the electronic band edges and SOC-driven valley splitting. Similarly, electric fields can alter the symmetry environment and redistribute charge density across the monolayer, leading to controllable changes in valley polarization. Recent studies demonstrated that strain tuned the band gap, spin filtering, valley splitting, Berry curvature, and anomalous Hall response in several 2D magnetic and ferrovalley systems~\cite{wang2025two,zeng2012valley,wang2024large,chang1996berry,xiao2010berry,liu2019valleytronics}. Thus, strain- and electric-field-controlled valley polarization in V-alloyed MoTe$_2$ may provide a practical route toward reconfigurable valleytronic devices.

In this work, we systematically investigated the structural stability, electronic properties, and valley polarization of the V-alloyed MoTe$_2$ monolayer, Mo$_{0.75}$V$_{0.25}$Te$_2$, using first-principles density functional theory (DFT) calculations. Here, the energetic and dynamical stability of the alloyed structure were examined through formation-energy and phonon-dispersion calculations. We first analyzed pristine MoTe$_2$ to establish the SOC-induced spin splitting and confirm the absence of valley polarization. We examined whether V alloying broke the valley degeneracy and produced a spontaneous valley polarization in the valence band. Furthermore, we analyzed the tunability of this valley polarization by applying electric fields along different crystal directions and biaxial compressive and tensile strain. The results and insights of these comprehensive analyses will be instrumental in designing V-alloyed MoTe\textsubscript{2}-based valleytronic devices. 

\section{Computational details}
We performed all first-principles calculations within the framework of DFT using the Quantum ESPRESSO package \cite{giannozzi2017advanced}. We carried out geometry optimization using the Perdew-Burke-Ernzerhof (PBE) generalized gradient approximation (GGA). We used the projector augmented wave (PAW) pseudopotentials to account for the interaction between the valence electrons and the core electrons \cite{hoque2022first}. We applied the Marzari-Vanderbilt-DeVita-Payne cold smearing method with a width of 0.01 Ry. For structural relaxation of MoTe\textsubscript{2} and Mo\textsubscript{0.75}V\textsubscript{0.25}Te\textsubscript{2}, we set the plane wave kinetic energy cutoff to 65 Ry, while the charge density cutoff was fixed at 650 Ry. We sampled the Brillouin zone using a $\Gamma$-centered Monkhorst-Pack grid of ($8\times 8 \times 1$) $k$-point for spin-polarized self-consistent field (SCF) calculations. For projected density of states (PDOS) calculations, we carried out the non-self-consistent field (NSCF) computations using a ($32\times 32 \times 1$) $k$-point grid. 

For structural relaxation, we set the convergence thresholds for the maximum force and the total energy below 1 meV\AA\textsuperscript{-1} and 10\textsuperscript{-5} eV, respectively. We performed non-collinear calculations including SOC to obtain accurate charge densities for band structure analysis. We computed the electronic band structures along the high-symmetry path $\Gamma$$\rightarrow$M+$\rightarrow$K+$\rightarrow$$\Gamma$ in the first Brillouin zone, with 20 $k$-points sampled between each pair of symmetry points. To examine the effects of SOC in detail, we carried out additional calculations along the extended path $\Gamma\rightarrow$ M+$\rightarrow$K+ $\rightarrow$$\Gamma$$\rightarrow$M--$\rightarrow$K-- $\rightarrow$$\Gamma$, again using 20 $k$-points between two adjacent $k$-path points.

\begin{figure*}[h]
\centering
\includegraphics[scale=0.37]{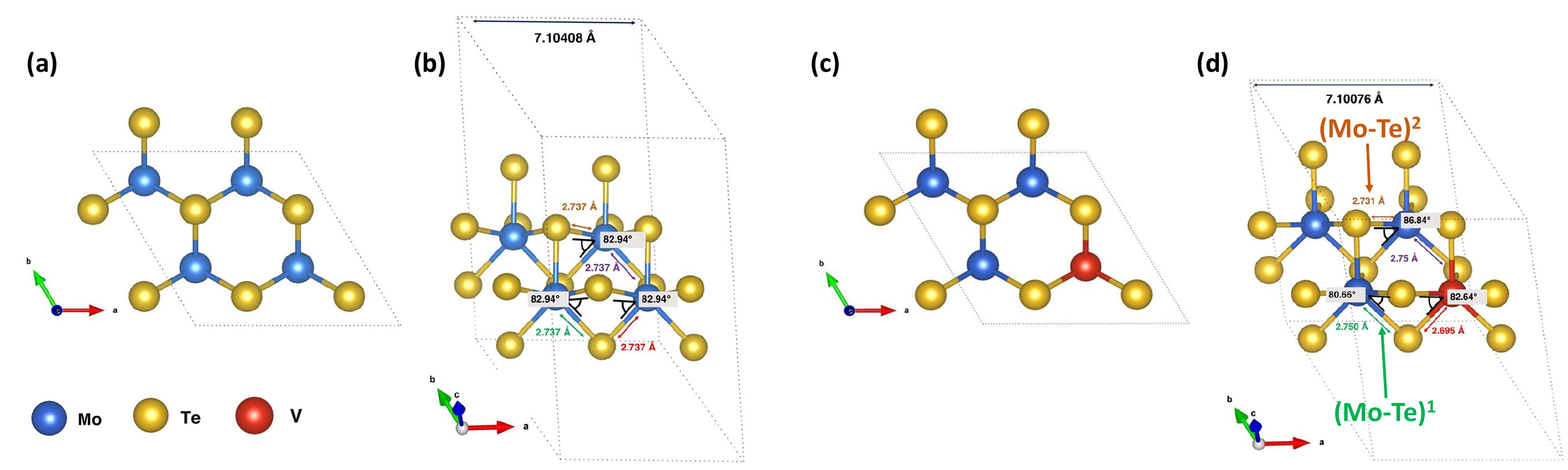}
 \caption{Structural details of (a) MoTe\textsubscript{2} top view, (b) MoTe\textsubscript{2} side view with lattice parameters, bond lengths, bond angles, (c) Mo\textsubscript{0.75}V\textsubscript{0.25}Te\textsubscript{2} top view, (d) Mo\textsubscript{0.75}V\textsubscript{0.25}Te\textsubscript{2} side view with lattice parameters, bond lengths, bond angles. Here, (Mo-Te)\textsuperscript{1} denotes a type-I bond between Mo and Te. Whereas (Mo-Te)\textsuperscript{2} denotes a type-II bond between Mo and Te.
}
  \label{Fig1}
\end{figure*}

\begin{table*}[t]
\centering
\small
\renewcommand{\arraystretch}{1.35}
\setlength{\tabcolsep}{6pt}
\caption{Crystal structure parameters.}
\label{tbl1}

\begin{tabularx}{\textwidth}{l c Y Y Y Y}
\toprule
\textbf{Crystal Structure} &
\shortstack{\textbf{Lattice}\\\textbf{Constant}\\\textbf{[a=b (\AA)]}} &
\multicolumn{2}{c}{\textbf{Bond Length (\AA)}} &
\multicolumn{2}{c}{\textbf{Bond Angle}} \\
\cmidrule(lr){3-4} \cmidrule(lr){5-6}
& & \textbf{Mo--Te} & \textbf{V--Te} &
\(\boldsymbol{\angle}\)\textbf{Te--Mo--Te} &
\(\boldsymbol{\angle}\)\textbf{Te--V--Te} \\
\midrule

MoTe$_2$
& 7.104
& 2.737
& ---
& 82.94$^\circ$
& --- \\

\shortstack[l]{Mo$_{0.75}$V$_{0.25}$Te$_2$\\(Mo$_3$VTe$_8$)}
& 7.101
& \begin{tabular}{@{}c@{}}2.750 (Mo-Te)\textsuperscript{1}\\2.731 (Mo-Te)\textsuperscript{2}\end{tabular}
& 2.695
& \begin{tabular}{@{}c@{}}86.84$^\circ$ (Mo-Te)\textsuperscript{1}\\80.66$^\circ$ (Mo-Te)\textsuperscript{2}\end{tabular}
& 82.64$^\circ$ \\

\bottomrule
\end{tabularx}
\end{table*}

We adopted a $\Gamma$-centered ($6\times 6 \times 1$) Monkhorst-Pack $k$-point mesh for both SCF and NSCF calculations, with SOC incorporated throughout. To enhance computational efficiency, we used a reduced $k$-point sampling while maintaining acceptable accuracy. We utilized fully relativistic PAW pseudopotentials, together with PBE from GGA, for band structure calculations including the effects of SOC in pristine MoTe$_2$ and V-alloyed Mo$_{0.75}$V$_{0.25}$Te$_2$. 
Unless otherwise specified, the magnetic moment of the V atom was aligned along the positive $z$-axis.
We introduced a vacuum layer of 15~\AA\ along the $c$-axis to eliminate spurious interactions between periodic images of adjacent monolayers. We carried out the phonon dispersion calculations using the Phonopy package to assess the dynamical stability of Mo\textsubscript{0.75}V\textsubscript{0.25}Te\textsubscript{2}. We generated a set of ($2 \times 2 \times 1$) supercells (72 in total) using the finite-displacement method. We calculated the force constants from SCF calculations employing an ($8\times 8 \times 1$) $k$-point mesh.

For the calculation of the energetic stability of the V alloyed MoTe\textsubscript{2} structure, we carried out the SCF calculations for Mo\textsubscript{0.75}V\textsubscript{0.25}Te\textsubscript{2}, MoTe\textsubscript{2} containing a single Mo vacancy, and an isolated V atom. We used the total energies obtained from these calculations to evaluate the formation energy of Mo\textsubscript{0.75}V\textsubscript{0.25}Te\textsubscript{2} system. The formation energy is expressed as

\begin{equation}
\mathrm{E}\textsubscript{formation} = \mathrm{E}\textsubscript{total}-\mathrm{E}\textsubscript{defect}-\mathrm{E}\textsubscript{Vanadium}
\label{eqn1}
\end{equation}
Where E\textsubscript{total} represents the total energy of Mo\textsubscript{0.75}V\textsubscript{0.25}Te\textsubscript{2}, E\textsubscript{defect} denotes the total energy of MoTe\textsubscript{2} with a single Mo vacancy, and E\textsubscript{Vanadium} corresponds to the energy of an isolated V atom.

To quantify the degree of valley polarization in the system, we defined an energy difference between the inequivalent valleys. The magnitude of valley polarization is given by
\begin{equation}
\Delta_{\mathrm{K\textsubscript{+}}\mathrm{K\textsubscript{--}}} = \mathrm{E}\textsubscript{K+}-\mathrm{E}\textsubscript{K--}
\label{eqn2}
\end{equation}
Where E\textsubscript{K+} and E\textsubscript{K--} correspond to the energies at the K+ and the K-- points
at the valence band maxima (VBM) or the conduction band minima (CBM), respectively.

To quantify spin splitting, let us denote the spin splittings as $\Delta$K\textsubscript{1} and $\Delta$ K$'$\textsubscript{1} at the K+ and the K-- valleys with the same sign created by SOC. On the other hand, SOC also produces spin splitting at the K+ and the K-- valleys with the opposite sign. We denoted these splittings
as -$\Delta$K\textsubscript{2} and $\Delta$K$'$\textsubscript{2}. We get,

\begin{equation}
\text{spin splitting at the K+}\text{ valley} = \Delta\mathrm{K\textsubscript{1}} - \Delta\mathrm{K\textsubscript{2}}
\label{eqn3}
\end{equation}
whereas,
\begin{equation}
\text{spin splitting at the K--}\text{ valley} = \Delta\mathrm{K}'\textsubscript{1} + \Delta\mathrm{K}'\textsubscript{2}
\label{eqn4}
\end{equation}

As discussed before, valley polarization can be further enhanced by applying external strain, either uniaxial or biaxial, to the material. The biaxial tensile strain is defined as
\begin{equation}
\varepsilon = \frac{a-a\textsubscript{0}}{a\textsubscript{0}},
\end{equation}
where $a\textsubscript{0}$ represents the lattice constant of unstrained \\ Mo\textsubscript{0.75}V\textsubscript{0.25}Te\textsubscript{2} and $a$ denotes the lattice constant of the strained system. Similarly, the biaxial compressive strain is expressed as

\begin{center}
\begin{equation}
\varepsilon = \frac{a\textsubscript{0}-a}{a\textsubscript{0}}.
\end{equation}
\end{center}


\section{Results and discussion}

\subsection{Structural evolution from MoTe$_2$ to V-alloyed MoTe$_2$ 
}

\begin{figure*}[h]
\centering
\includegraphics[scale=0.25]
{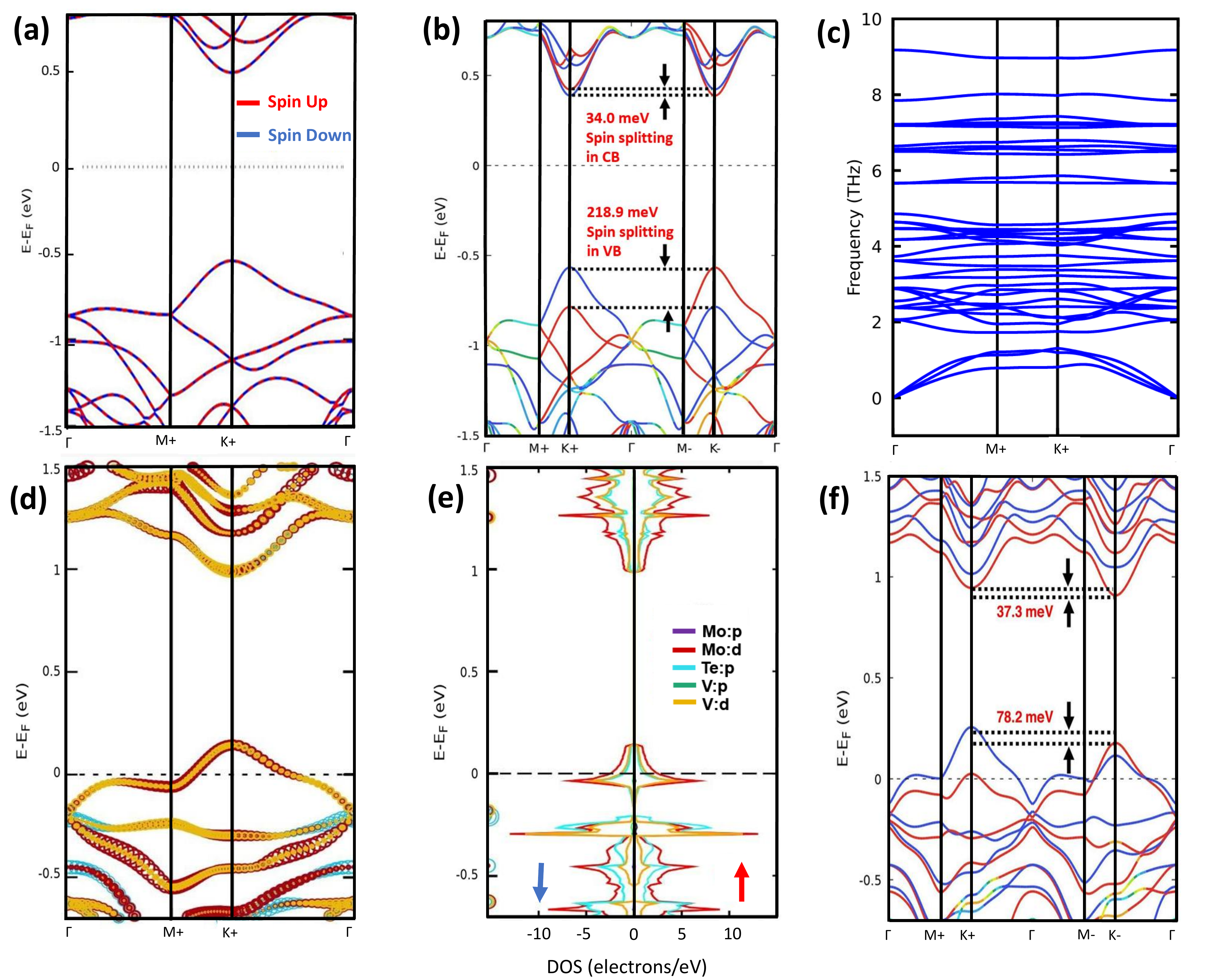}
 \caption{Spin polarized band structure of pristine MoTe\textsubscript{2} (a) without SOC, (b) with SOC. Red color indicates spin-up and blue color
indicates spin-down band. CB means conduction band, and VB means valence band, (c) phonon band dispersion curve of Mo\textsubscript{0.75}V\textsubscript{0.25}Te\textsubscript{2}, (d) projected band structure of Mo\textsubscript{0.75}V\textsubscript{0.25}Te\textsubscript{2}, (e) projected density of states of Mo\textsubscript{0.75}V\textsubscript{0.25}Te\textsubscript{2}, and 
(f) spin polarized band structure of Mo\textsubscript{0.75}V\textsubscript{0.25}Te\textsubscript{2} with SOC. 
}
  \label{Fig2}
\end{figure*}

We incorporated a V atom into the  MoTe\textsubscript{2} lattice to induce magnetism. \autoref{Fig1} illustrates the optimized structure of the pristine monolayer MoTe\textsubscript{2}, including its bond lengths, bond angles, and lattice parameters. We constructed a ($2\times 2\times 1$) supercell from the primitive unit cell of pristine MoTe\textsubscript{2}. The primitive unit cell contained three atoms, comprising one Mo atom and two Te atoms. Therefore, the ($2\times 2\times 1$) supercell consisted of four Mo atoms and eight Te atoms.
We substituted one of the four Mo atoms in the ($2\times 2\times 1$) supercell with a V atom, yielding the composition 
Mo\textsubscript{0.75}V\textsubscript{0.25}Te\textsubscript{2} (or equivalently Mo\textsubscript{3}VTe\textsubscript{8}). We found the lattice parameter of the pristine MoTe\textsubscript{2} to be 3.552 \AA, which is in good agreement with previously reported values in the literature \cite{huang2016controlling}. The calculated Mo--Te bond length and $\angle \mathrm{Te-Mo-Te}$ bond angle were 2.737 \AA\ and 82.94$^\circ$, respectively. The introduction of substitutional V alloying modified the structural parameters of the system, including 
the lattice constant, bond lengths, and bond angles. We obtained a lattice parameter
of 7.101 \AA\ for Mo\textsubscript{0.75}V\textsubscript{0.25}Te\textsubscript{2}. The optimized V-Te bond length was found to be 2.695 ~\AA, while two different types of Mo-Te bonds were identified in the optimized structure.
We classified the Mo–Te bond closest to the V alloying site as a type-I bond labeled (Mo-Te)\textsuperscript{1}.
In addition, \autoref{Fig1} also shows another distinct Mo–Te bond, labeled (Mo–Te)\textsuperscript{2}, which corresponds to a type-II bond and is located further from the V atom within the lattice. The length of the type-I bond increased to 2.750 \AA\ which is demonstrated in \autoref{Fig1}. However, the length of the type-II bond decreased to 2.750 \AA. We obtained a bond angle $\angle \mathrm{Te-V-Te}$ of $82.64^\circ$, while the bond angle of $\angle \mathrm{Te-Mo-Te}$ in the vicinity of the V site (type-I region) decreased to $80.66^\circ$. In contrast, Mo atoms located farther from the V site (type-II region) exhibited a larger $\angle \mathrm{Te-Mo-Te}$ bond angle of $86.84^\circ$. Additional structural details are provided in Figs. S1-S2 of the Supplementary Material.

\begin{figure*}[h]
\centering
\includegraphics[width=\textwidth]
{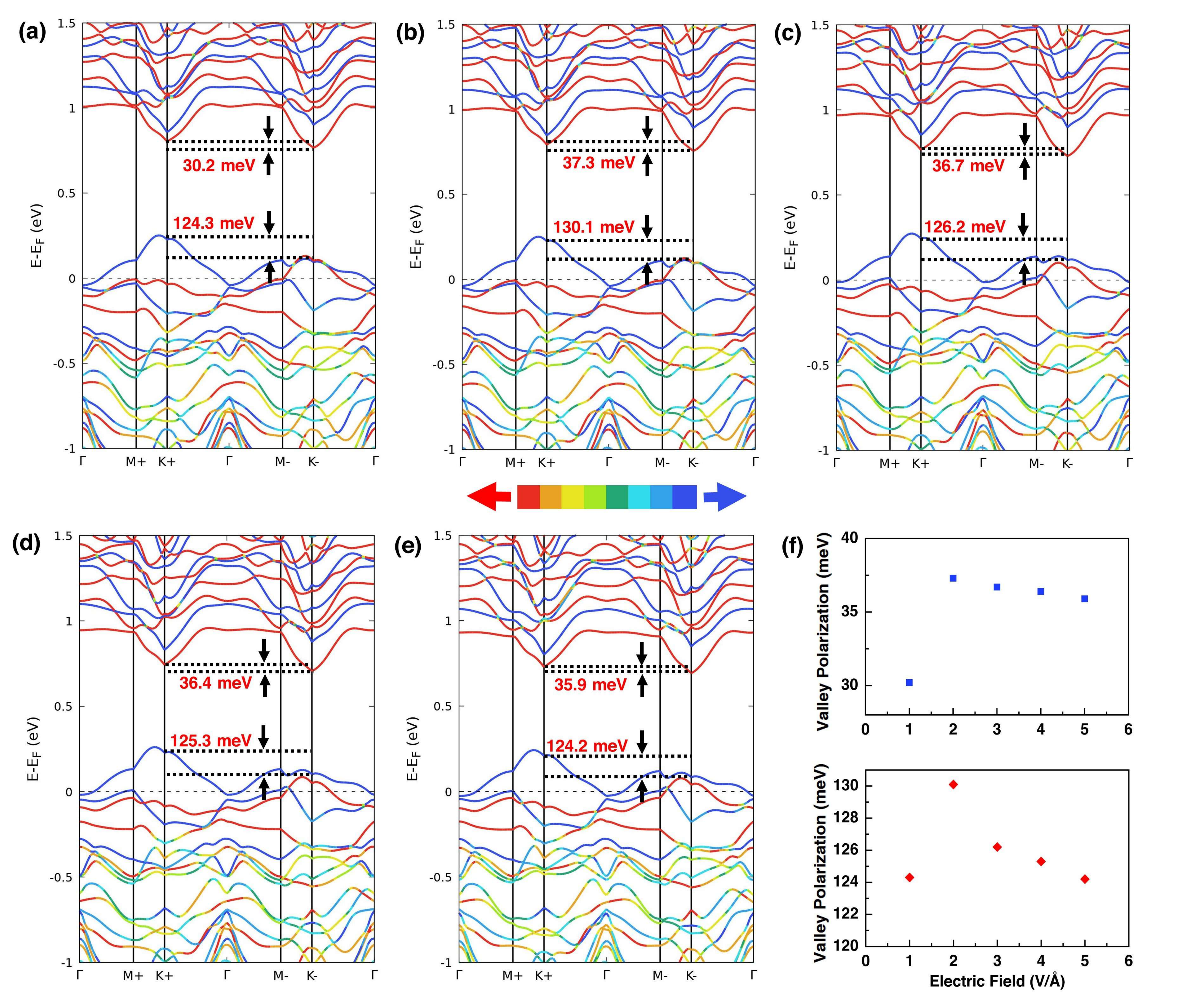}
\caption{Electric field applied along the crystal a axis on Mo\textsubscript{0.75}V\textsubscript{0.25}Te\textsubscript{2}. The band structures
with SOC for (a) 0.1 \VperA, (b) 0.2 \VperA, (c) 0.3 \VperA, (d) 0.4 \VperA, (e) 0.5 \VperA, (f) Plot of valley
splitting vs applied electric field. The top blue curve denotes the conduction band valley splitting, and the bottom red curve denotes the valence band valley splitting.}
  \label{Fig5}
\end{figure*}


\subsection{Electronic band engineering via transition-metal alloying} 
We investigated the electronic properties of pristine MoTe\textsubscript{2} to validate our results against previously reported literature. The spin-polarized band structure of pristine MoTe\textsubscript{2}, as shown in \autoref{Fig2} (b), revealed no spin splitting in the conduction band and the valence band. This observation indicates that the up-spin and the down-spin bands are degenerate in pristine MoTe\textsubscript{2}. The calculated band gap of pristine MoTe\textsubscript{2} was found to be 1.09 eV, which is consistent with the literature \cite{ruppert2014optical}. Moreover, we noticed that the Fermi level was right between the conduction band and the valence band. Since
the electronic band gap is 1.09 eV, pristine MoTe\textsubscript{2} is a semiconductor material. Subsequently, we studied the band structure of pristine MoTe\textsubscript{2} with the effect of SOC. 
Interestingly, the band gap decreased to 0.95 eV after including SOC. 
In addition, we identified spin
splitting in both the conduction band and the valence band at the K+ and K-- points, with the valence band exhibiting a larger magnitude of 
spin splitting compared to that of the conduction band. In the conduction band, we determined a spin splitting of 34 meV, while a spin
splitting of 218.9 meV was observed in the valence band according to \autoref{eqn3} and \autoref{eqn4} respectively. However, the bands remained degenerate at the K+ and K-- points, indicating the absence of valley polarization in pristine
MoTe\textsubscript{2}. A similar analysis of VTe\textsubscript{2} is presented in Figs. S5-S6 of the Supplementary Material. To analyze the electronic properties of Mo\textsubscript{0.75}V\textsubscript{0.25}Te\textsubscript{2}, we calculated the projected band
structure. From the projected band structure shown in \autoref{Fig2} (d), we observed that the CBM and the VBM occured at the same $k$-point in the momentum space. 

\begin{figure*}[h]
\centering
\includegraphics[width=\textwidth]
{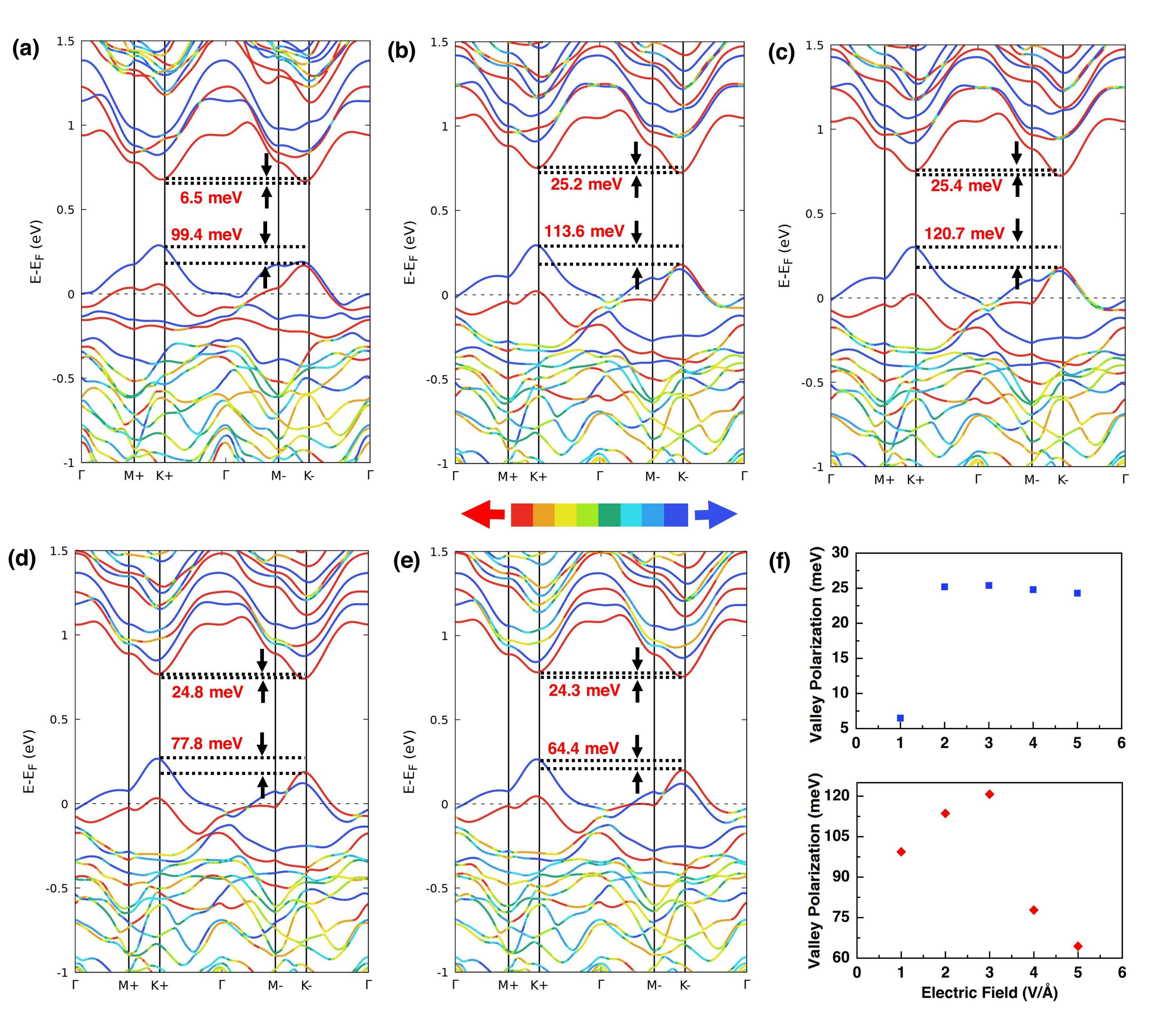}
\caption{Electric field applied along the crystal b axis on Mo\textsubscript{0.75}V\textsubscript{0.25}Te\textsubscript{2}. The band structures
with SOC for (a) 0.1 \VperA, (b) 0.2 \VperA, (c) 0.3 \VperA, (d) 0.4 \VperA, (e) 0.5 \VperA, (f) Plot of valley
splitting vs applied electric field. The top blue curve denotes the conduction band valley splitting, and the bottom red curve denotes the valence band valley splitting.}
  \label{Fig6}
\end{figure*}

The CBM is dominated by the d orbitals of both Mo and V atoms, with a higher contribution from the V d orbitals compared to those of Mo. The CBM
consists primarily of the d\textsubscript{z\textsuperscript{2}} orbital of V (32.9\%) and the d\textsubscript{z\textsuperscript{2}} orbital of Mo (24\%). Additional contributions arise from the d\textsubscript{x\textsuperscript{2}-y\textsuperscript{2}} (9.9\%), d\textsubscript{xy} (10\%),
and s (4.5\%) orbitals of Mo, as well as the p\textsubscript{x} (6.3\%), p\textsubscript{y} (7.4\%), p\textsubscript{z} (1.5\%) orbitals of Te along with a negligible contribution from the
s orbital of V (0.1\%). In contrast, the VBM also originates mainly from the d
orbitals of Mo and V, but with an opposite trend in orbital dominance, where Mo contributed more strongly than V. The VBM
consists primarily of the d\textsubscript{xy} (13.6\%), and d\textsubscript{x\textsuperscript{2}-y\textsuperscript{2}} (13.6\%) orbitals of V, together with the d\textsubscript{xy} (27.3\%), and d\textsubscript{x\textsuperscript{2}-y\textsuperscript{2}} (27.3\%) orbitals of Mo. Additional contributions originate from the d\textsubscript{z\textsuperscript{2}} (0.3\%) orbital of Mo: the p\textsubscript{x} (7.2\%), and p\textsubscript{y} (7.2\%) orbitals of Te. From the
analysis of the projected band structure, we clearly established that the valley polarization phenomena would be strongly governed by the d orbitals of the transitional metal atoms.

\subsection{Energetic and dynamical stability of the alloyed monolayer} 
We examined the stability of the Mo\textsubscript{0.75}V\textsubscript{0.25}Te\textsubscript{2} alloy from both energetic and dynamical perspectives. To evaluate its energetic stability, we performed SCF calculations for three reference systems: the V-alloyed MoTe\textsubscript{2} monolayer, MoTe\textsubscript{2} containing a single Mo vacancy, and an isolated V atom. Following this, we obtained the total energies from these SCF calculations to evaluate the formation energy of the Mo\textsubscript{0.75}V\textsubscript{0.25}Te\textsubscript{2} alloy using \autoref{eqn1}. The calculated formation energy was found to be \(-0.84\) eV, indicating that the substitution of a Mo atom by a V atom is energetically favorable. 
\begin{figure*}[h]
\centering
\includegraphics[width=\textwidth]
{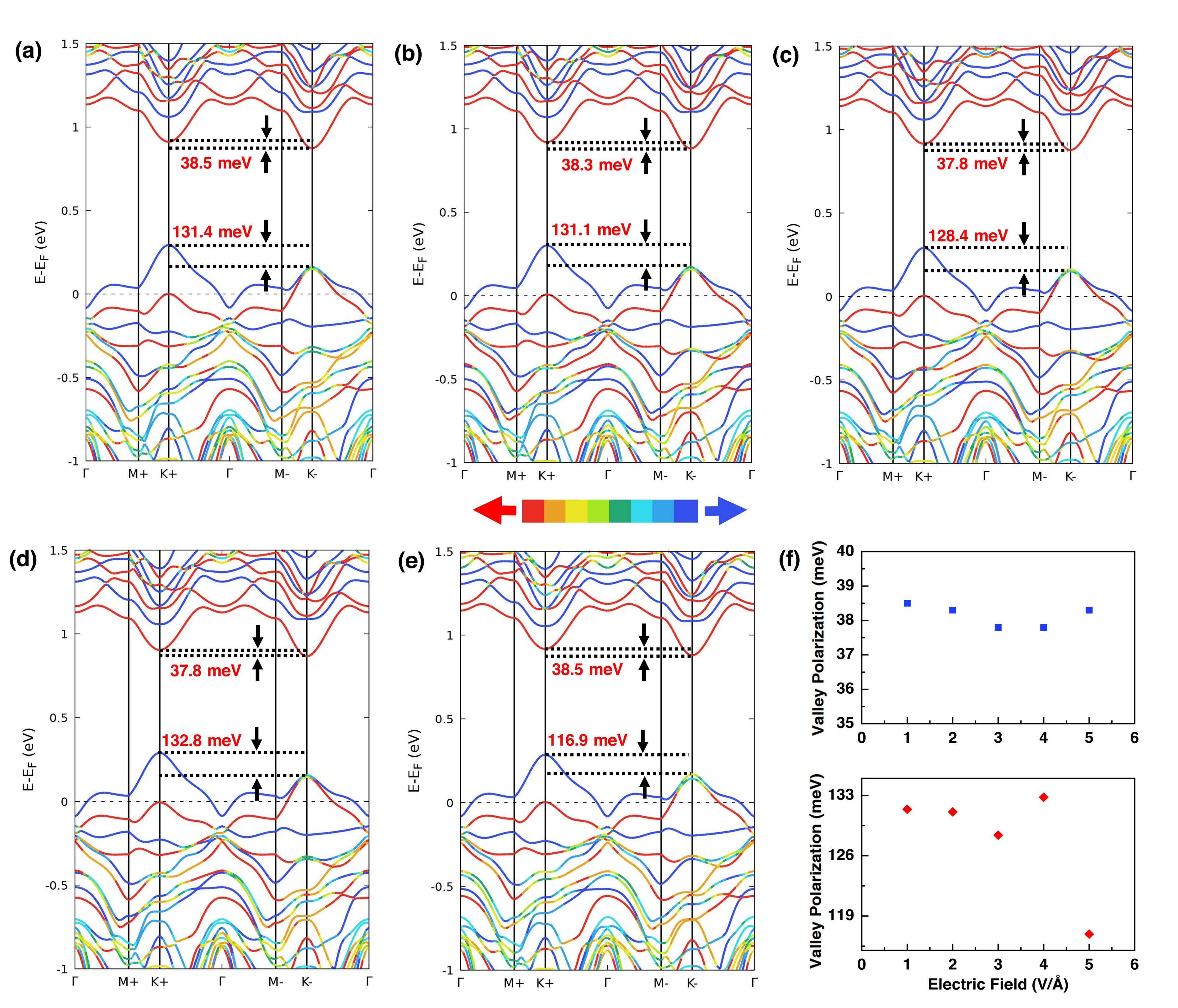}
\caption{Electric field applied along the crystal c axis on Mo\textsubscript{0.75}V\textsubscript{0.25}Te\textsubscript{2}. The band structures
with SOC for (a) 0.1 \VperA, (b) 0.2 \VperA, (c) 0.3 \VperA, (d) 0.4 \VperA, (e) 0.5 \VperA, (f) Plot of valley
splitting vs applied electric field. The top blue curve denotes the conduction band valley splitting, and the bottom red curve denotes the valence band valley splitting.}
  \label{Fig7}
\end{figure*}
The negative value of the formation energy suggests that the Mo\textsubscript{0.75}V\textsubscript{0.25}Te\textsubscript{2} alloy can form spontaneously under suitable conditions and possesses good energetic stability.

To further confirm the stability of the Mo\textsubscript{0.75}V\textsubscript{0.25}Te\textsubscript{2} alloy, we investigated it's dynamic stability via phonon dispersion calculations. The phonon band structure of\\ Mo\textsubscript{0.75}V\textsubscript{0.25}Te\textsubscript{2}, shown in \autoref{Fig2} (c), does not exhibit any negative (imaginary) phonon frequencies throughout the Brillouin zone. The absence of imaginary modes indicates that the optimized Mo\textsubscript{0.75}V\textsubscript{0.25}Te\textsubscript{2} alloys does not undergo lattice instability against small atomic displacements. The phonon dispersion result confirms the dynamic stability of Mo\textsubscript{0.75}V\textsubscript{0.25}Te\textsubscript{2}. The dynamic stability of pristine MoTe\textsubscript{2} and pristine VTe\textsubscript{2} is discussed in sections S2-S3 of the Supplementary Material.

\subsection{Tunable valley polarization in Mo\textsubscript{0.75}V\textsubscript{0.25}Te\textsubscript{2}}

In two-dimensional hexagonal materials, breaking spatial inversion symmetry induces a valley-contrasting Berry curvature, $\Omega(\mathbf{k}) = -\Omega(-\mathbf{k})$, and heavy atomic SOC leads to spin-valley locking at the degenerate K+ and K-- valleys, which allows for selective optical initialization via circular dichroism and drives the valley Hall effect. Introducing mechanical strain alters local atomic hopping parameters without breaking time-reversal symmetry, which acts mathematically as a deformation-induced gauge field shifting the reciprocal valleys in opposite directions ($\mathbf{A}_{\mathrm{K+}} = -\mathbf{A}_{\mathrm{K--}}$) to generate a pseudo-magnetic field that establishes valley-filtering transport barriers. 
\begin{figure*}[h]
\centering
\includegraphics[width=\textwidth]
{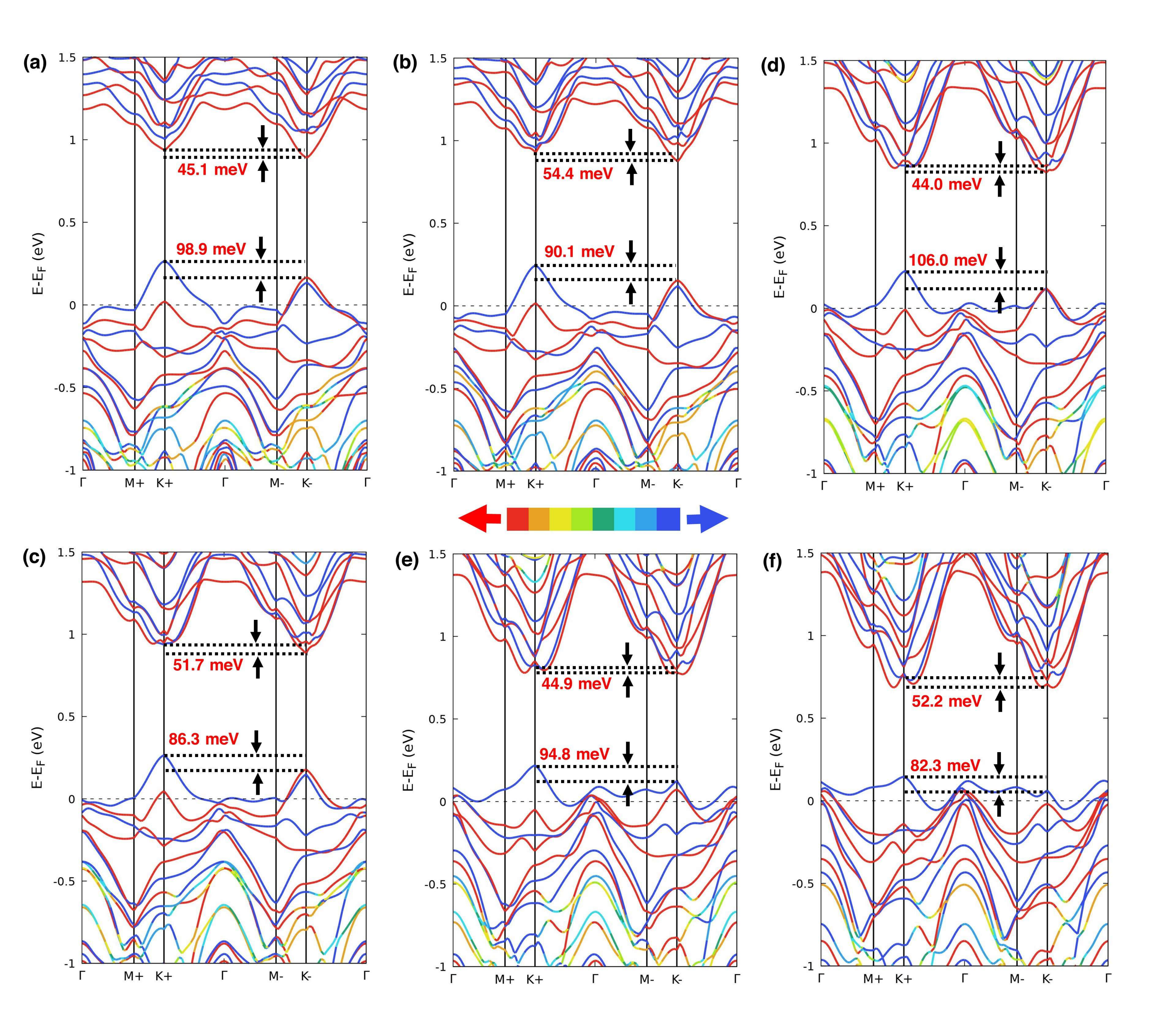}
\caption{Compressive biaxial strain applied along the crystal ab plane. The band structures with
SOC for (a) 1\%, (b) 2\%, (c) 3\%, (d) 4\%, (e) 5\%, (f) 6\% compressive strain.}
  \label{Fig8}
\end{figure*}
Meanwhile, an external electric field governs valley mechanics depending on its orientation: an out-of-plane field ($\mathbf{E}_z$) breaks layer-inversion symmetry in multi-layer structures to electrostatic-gatedly open a tunable energy gap, whereas an in-plane electric field ($\mathbf{E}_{x,y}$) induces a transport momentum drift that, when combined with strain-induced orbital distortion, breaks the structural equivalence of electron-hole wavefunction overlaps between the K+ and K-- valleys to achieve pure electrical control over valley polarization.

\subsubsection{Intrinsic valley polarization of pristine MoTe\textsubscript{2}}
We studied the band structure of the pristine MoTe\textsubscript{2} with the effect of SOC depicted in \autoref{Fig2} (b). Interestingly, the inclusion of SOC reduced the band gap to 0.95 eV. Furthermore, we identified spin splitting in both the conduction band and the valence band at the K+ and K− points, with the valence band exhibiting a larger splitting magnitude explained before. Therefore, pristine MoTe\textsubscript{2} did not exhibit valley polarization.

\subsubsection{Valley engineering in Mo\textsubscript{0.75}V\textsubscript{0.25}Te\textsubscript{2}(Mo\textsubscript{3}VTe\textsubscript{8})}
Previously, we introduced the expression given in \autoref{eqn2} to measure the degree of valley polarization, along with the expressions presented in \autoref{eqn3} and \autoref{eqn4}
to evaluate the degree of spin splitting at the K+ valley and the K-- valley, respectively. \autoref{Fig2} (f) shows that,  Mo\textsubscript{0.75}V\textsubscript{0.25}Te\textsubscript{2} has K+ and K-- valleys which are clearly distinguishable and energetically non-degenerate. We observed that the K-- valley lies at a slightly lower energy than the K+ valley. 
Therefore, \autoref{Fig2} (f) clearly depicts valley polarization both in the conduction band and in the valence band. In the valence
band, the valley polarization is 78.2 meV, whereas in the conduction band it is 37.3 meV according to \autoref{eqn2}. 
The underlying mechanism of this valley polarization in Mo\textsubscript{0.75}V\textsubscript{0.25}Te\textsubscript{2} can be attributed to the combined
effects of transition-metal alloying and SOC. Regarding the first mechanism, transition-metal alloying
breaks time reversal symmetry, which leads to

\begin{figure*}[h]
\centering
\includegraphics[width=\textwidth]
{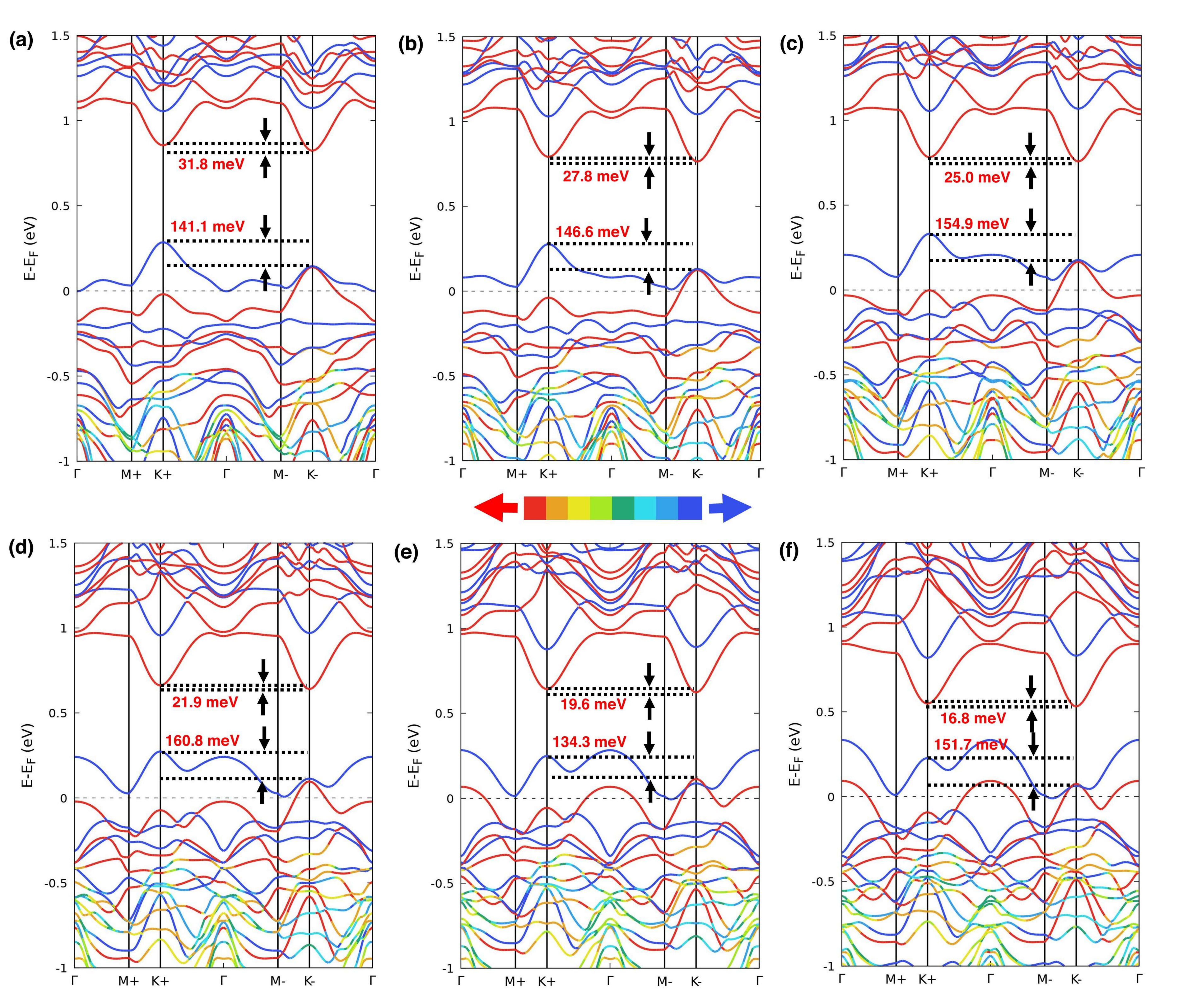}
\caption{Tensile biaxial strain applied along the crystal ab plane. The band structures with SOC for (a) 1\%, (b) 2\%, (c) 3\%, (d) 4\%, (e) 5\%, (f) 6\% tensile strain.}
  \label{Fig9}
\end{figure*}

identical spin splitting at the K+ and the K-- valleys. 
An interesting consequence of this valley polarization is the emergence of an anomalous Hall effect (AHE) in two-dimensional materials, as approved by the previous study \cite{tong2016concepts}.

\subsubsection{Electric-field-controlled modulation of valley polarization in Mo\textsubscript{0.75}V\textsubscript{0.25}Te\textsubscript{2}(Mo\textsubscript{3}VTe\textsubscript{8})}
The main goal of this paper revolves around the fact that how to possibly enhance this valley polarization. We primarily employed
two approaches: (1) application of an external electric field along different crystallographic directions, and (2) application of biaxial strain to the crystal structure. To enhance the valley polarization already existed in the Mo\textsubscript{0.75}V\textsubscript{0.25}Te\textsubscript{2} structure,
we applied an electric field along the crystal a, b, c direction. The magnitude of
the applied electric field was varied from 0.1 to 0.5 \VperA.
When we applied an electric field along the crystallographic a-axis, we obtained notable changes in the electronic properties. As shown in \autoref{Fig5}, valley polarization increased significantly. The energy degeneracy in both the conduction band and the valence band
at the K+ and K-- points was strongly lifted. 
\begin{figure*}[h]
\centering
\includegraphics[width=\textwidth]
{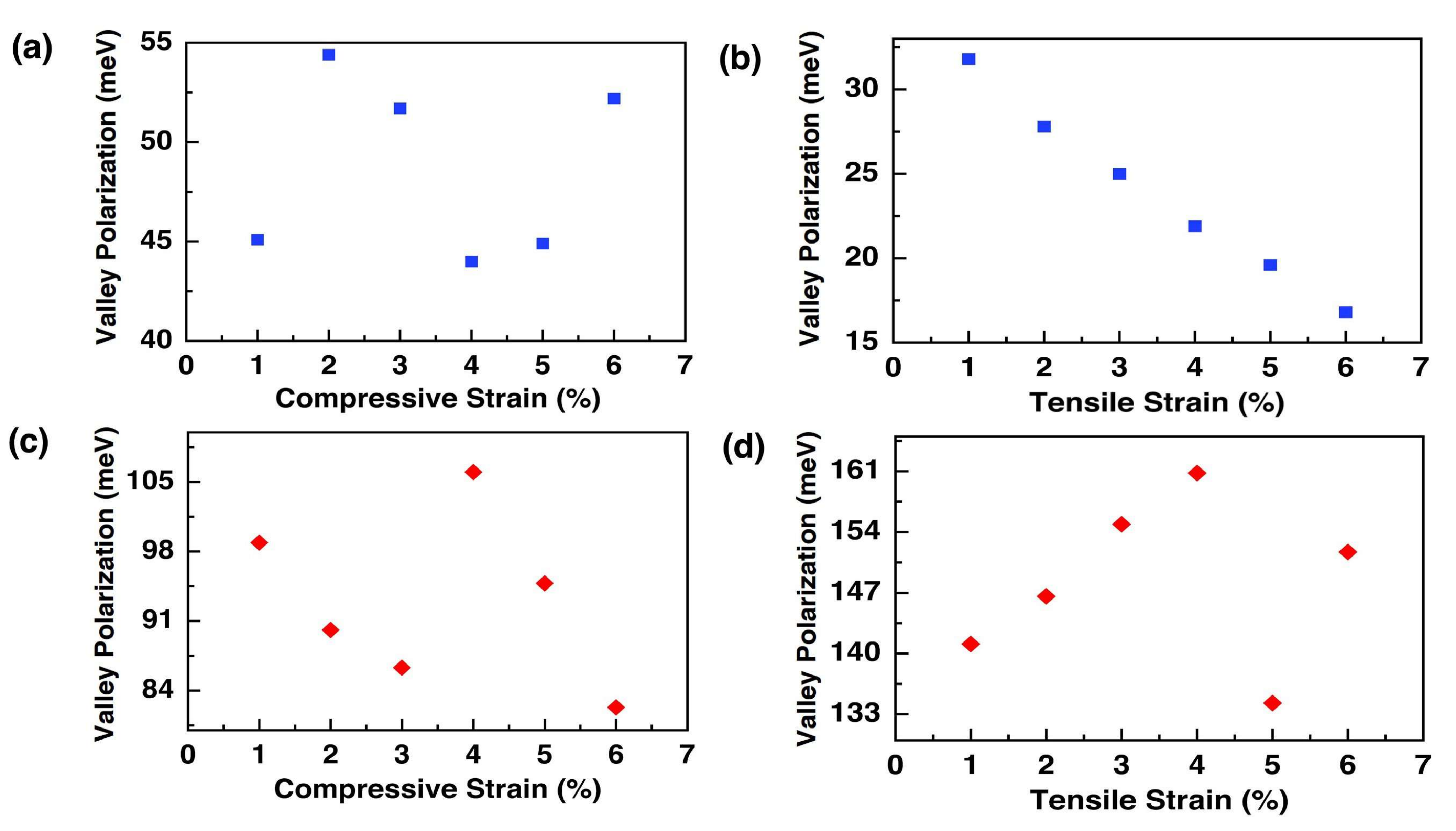}
\caption{Valley polarization vs applied biaxial strain for Mo\textsubscript{0.75}V\textsubscript{0.25}Te\textsubscript{2} (a) Valley polarization in the conduction band vs applied compressive strain, (b) Valley polarization in the conduction band vs applied tensile strain, (c) Valley polarization in the valence band vs applied compressive strain, and (d) Valley polarization in the valence band vs applied tensile strain}
  \label{Fig10}
\end{figure*}
The valley polarizations in the valence band was found to be 124.3, 130.1, 126.2, 125.3, and 124.2 meV 
for applied electric fields of magnitude 0.1, 0.2,
0.3, 0.4, and 0.5 \VperA, respectively. In the conduction band, the corresponding valley polarization values were 30.2, 37.3, 36.7, 36.4, and 35.9 meV for applied electric
fields of 0.1, 0.2, 0.3, 0.4, and 0.5 \VperA, respectively. We obtained the maximum valley polarization for an applied electric field of 0.2 V/\AA, where the energy difference between the VBM at the K+ and K-- points reached 130.1 meV, indicating a giant valley polarization.
The conduction band also exhibited a valley polarization of 37.3 meV under an applied planar electric field of 0.2 V/\AA\ along the crystallographic a-axis.

When we applied an electric field along the crystallographic b-axis, the valley polarization
changed significantly. As shown in \autoref{Fig6}, the valley polarization
decreased significantly compared to the case where the electric field
was applied along the crystallographic a-axis. Although, the splitting in the valence band in both the K+ and K-- points increased compared to the
relaxed Mo\textsubscript{0.75}V\textsubscript{0.25}Te\textsubscript{2} structure. The valley polarizations in the valence
band was found to be 99.4, 113.6, 120.7, 77.8, and 64.4 meV for applied electric
fields of magnitude 0.1, 0.2, 0.3, 0.4, and 0.5 \VperA, respectively. In
the conduction band, the valley polarization values were 6.5, 25.2, 25.4,
24.8, and 24.3 meV for applied electric fields of 0.1, 0.2, 0.3, 0.4,
and 0.5 \VperA, respectively. The maximum valley polarization was obtained at an applied field of 0.3 V/\AA, where the energy difference at the VBM between the K+ and K-- points reached 120.7 meV, indicating a giant valley polarization.
An interesting observation is that, under the applied planar electric field, the
valley polarization increased with increasing magnitude of the electric field and
subsequently began to decrease. This phenomenon is evident
from the comparison plot of valley polarization vs electric field of \autoref{Fig5}(f) and
\autoref{Fig6}(f). 


To further investigate the characteristics of valley polarization in Mo\textsubscript{0.75}V\textsubscript{0.25}Te\textsubscript{2},
we applied a transverse electric field along the crystallographic c-axis, perpendicular to the plane of  Mo\textsubscript{0.75}V\textsubscript{0.25}Te\textsubscript{2}. We obtained notable changes in the electronic properties under this condition. As shown in \autoref{Fig7}, the valley polarization increased significantly. At the K-- point in the momentum space, the spin-up and spin-down bands overlapped, indicating the presence of pseudospin character in the valence band at that momentum point. In contrast, a pronounced splitting between the VBM
at the K+ and K-- points confirmed strong valley polarization. The valley polarizations in the valence band was found to be 131.4,
131.1, 128.4, 132.8, and 116.9 meV for applied electric fields of magnitude 0.1, 0.2, 0.3, 0.4, and 0.5 \VperA, respectively. In the conduction band,
the corresponding values were 38.5, 38.3, 37.8, 37.8, and 38.5
meV for applied electric fields of 0.1, 0.2, 0.3, 0.4, and 0.5 \VperA, respectively. The maximum valley polarization was obtained at an applied field of 0.4 V/\AA, where
the VBM energy difference between the K+ and K-- points reached to 132.8 meV, which is a giant
valley polarization. Another noteworthy observation is that the valley polarizations both in
the conduction band and the valence band remained nearly constant with increasing transverse electric strength.The observed valley polarization arises from the combined effects of strong SOC inherent in MoTe\textsubscript{2} and the effective exchange  field (Zeeman-field) introduced by V substitutional alloying.

\subsection{Strain-induced valley physics in Mo\textsubscript{0.75}V\textsubscript{0.25}Te\textsubscript{2}(Mo\textsubscript{3}VTe\textsubscript{8})}


As illustrated in \autoref{Fig8}, the application of biaxial compressive strain significantly enhanced the valley polarization. To investigate the strain-dependent electronic behavior, we applied biaxial compressive strains ranging from 1 to 6\% along the crystallographic ab-plane and analyzed the corresponding SOC induced band structures. We obtained a pronounced valley polarization in both the conduction band and the valence band between the K+ and K-- points. The valley polarizations in
the valence band was calculated to be 98.9, 90.1, 86.3, 106.0, 94.8, and 82.3 meV for compressive strains of 1, 2, 3, 4, 5, and 6\%, respectively. In the conduction band, the valley polarization values were 45.1, 54.4, 51.7, 44.0, 44.9, and 52.2 meV for compressive strains of 1, 2, 3, 4, 5, and 6\%, respectively. An interesting observation
is that biaxial compressive strain induced substantially larger valley polarization in the conduction band to previously investigated  electric-field-modulated cases. We obtained the maximum valley polarization by applying 4\% compressive strain, where the energy difference between the VBM at the K+ and K-- points reached to 106.0 meV. In contrast, the largest valley
polarization in the conduction band was found to be 54.4 meV under biaxial compressive strain of 2\%. Both values indicate the emergence of giant valley polarizations in the strained system.


When we applied biaxial tensile strain along the crystallographic ab-plane, we obtained the most pronounced enhancement of valley polarization among all the strategies investigated in this work. As illustrated in \autoref{Fig9}, the resulting exceeded the values obtained under electric fields and compressive strain. We applied 6 different biaxial tensile strains ranging from 1 to 6\% and analyzed the band structures with SOC. The calculated valley polarizations in the valence band were 141.1, 146.6, 154.9, 160.8, 134.3, and 151.7 meV for 1,
2, 3, 4, 5, and 6\% tensile strains, respectively. In the conduction band,
the valley polarization values were 31.8, 27.8, 25, 21.9, 19.6,
and 16.8 meV for tensile strains of 1,
2, 3, 4, 5, and 6\%, respectively. A particularly more interesting trend was that the conduction band valley polarization decreased progressively with increased tensile strain. The maximum valley polarization in the valence band was achieved at 4\% tensile strain, where the energy difference between the VBM at the K+ and K-- points reached 160.8 meV.
representing an exceptionally large valley polarization. In contrast, the largest
valley polarization in the conduction band was found to be 31.8 meV at a biaxial tensile strain of 1\%.
\begin{figure*}[h]
    \centering
    \includegraphics[width=\textwidth]{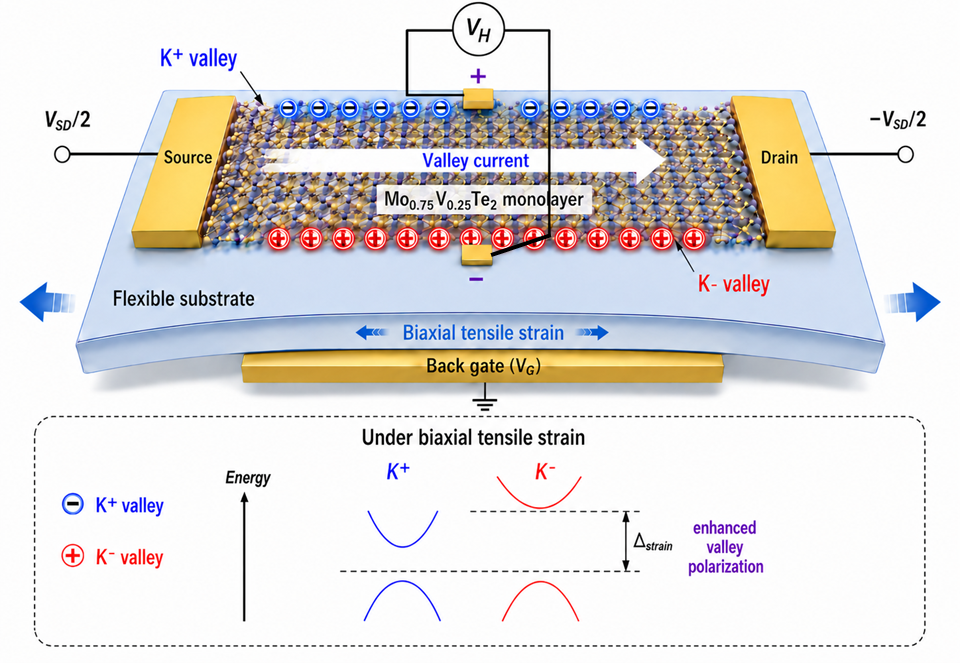}
    \caption{
    Schematic illustration of a strain-controlled valleytronic device based on
    monolayer \(\mathrm{Mo}_{0.75}\mathrm{V}_{0.25}\mathrm{Te}_{2}\).
    The alloyed monolayer is placed on a flexible substrate with source and
    drain electrodes, a back gate, and transverse Hall voltage probes. Under
    biaxial tensile strain, the energy separation between the K+ and
    K-- valleys increases, resulting in enhanced valley polarization and
    a measurable voltage.
    }
    \label{fig:application}
\end{figure*}
We plotted the valley polarization as a function of applied strain in \autoref{Fig10}. Based on the results presented in \autoref{Fig5}, \autoref{Fig6}, \autoref{Fig7}, \autoref{Fig8}, and \autoref{Fig9}, several important observations can be made. The enhancement of valley polarization can be attributed primarily to two factors: SOC and transition-metal alloying. When SOC was considered, opposite spin states dominated the CBM and the VBM. Upon introducing the transitional metal (V) atom into the pristine MoTe\textsubscript{2}, an effective Zeeman field appeared, the valley polarization was dominated by the Zeeman interaction. The opposite spins in the CBM and VBM occupied
non-degenerate energy levels at the K+ and K-- valleys because of the Zeeman interaction. We noticed another thing that, the K+ valley is higher than the K-- valley in both the CBM and VBM.

\subsection{Application: strain-controlled valleytronic device based on Mo\textsubscript{0.75}V\textsubscript{0.25}Te\textsubscript{2}}

The large and tunable valley polarization obtained in 
\(\mathrm{Mo}_{0.75}\mathrm{V}_{0.25}\mathrm{Te}_{2}\) suggests its potential
application as an active channel material for strain-controlled valleytronic
devices. A possible device concept is illustrated in \autoref{fig:application},
where a monolayer \(\mathrm{Mo}_{0.75}\mathrm{V}_{0.25}\mathrm{Te}_{2}\)
channel is placed on a flexible substrate and connected to source and drain
electrodes. A longitudinal source--drain bias, \(V_{SD}\), drives carrier
transport through the channel, while a bacK--gate voltage, \(V_G\), can be used
to tune the carrier concentration and position the Fermi level near the desired
valley-polarized band edge. A transverse Hall voltage, \(V_H\), can then be
detected across the lateral probes due to the imbalance between the
inequivalent K+ and K-- valleys.

The operating principle of the proposed device relies on the valley-dependent
electronic structure of V-alloyed \(\mathrm{MoTe}_{2}\). In pristine
\(\mathrm{MoTe}_{2}\), SOC produces spin splitting, but unequal valleys remain energetically degenerate due to the
preserved time-reversal symmetry. After V alloying, magnetic exchange
interaction breaks time-reversal symmetry, and the combined effect of exchange
interaction and SOC lifts the valley degeneracy. As a result,
\(\mathrm{Mo}_{0.75}\mathrm{V}_{0.25}\mathrm{Te}_{2}\) 
exhibits spontaneous
valley polarization of \(37.3~\mathrm{meV}\) in the conduction band and
\(78.2~\mathrm{meV}\) in the valence band. This intrinsic valley polarization
provides the fundamental requirement for valley-selective carrier transport.

The proposed device becomes particularly attractive when the monolayer is
integrated with a flexible substrate. Applying biaxial tensile strain modifies
the lattice constant, bond length, orbital hybridization, and exchange--SOC
coupling of the alloyed monolayer. According to the calculated results, tensile
strain produced the strongest enhancement of valley polarization, increasing
the valence-band valley splitting up to \(160.8~\mathrm{meV}\) at \(4\%\)
tensile strain. This value is significantly larger than the spontaneous valley
splitting of the relaxed structure, indicating that mechanical deformation can
serve as an efficient external control knob for valley polarization. By stretching or bending the flexible substrate, the valley splitting can be
dynamically modulated, enabling a strain-controlled valleytronic sensor or
switch.

In the device configuration shown in \autoref{fig:application}, tensile strain
increases the energy separation between the K+ and K-- valleys, as
illustrated in the lower panel of the figure. When a longitudinal bias is
applied, carriers originating from the two valleys acquire opposite velocities due to their valley-contrasting behavior. Since the two
valleys are no longer energetically equivalent, one valley contributes more
strongly to transport than the other, producing a net transverse Hall signal.
The magnitude of this Hall voltage is expected to depend on the degree of
valley polarization and, consequently, on the applied strain. This provides a
direct mechanism for converting mechanical deformation into an electrical
valleytronic signal.

The back gate further improves the functionality of the device by enabling
electrostatic control of the carrier density. By tuning \(V_G\), the Fermi
level can be positioned near either the valence band or conduction band
valley-polarized states. Since the present calculations showed that the valence
band gives the largest valley splitting under tensile strain, hole-mediated
operation may be particularly favorable for achieving a strong valley Hall
response. On the other hand, the conduction band also showed sizable tunability,
with the conduction band valley splitting reaching
\(54.4~\mathrm{meV}\) under \(2\%\) biaxial compressive strain. This suggests
that both electron- and hole-based valleytronic operation modes may be possible
depending on the device configuration.

\subsection{Comparative analysis}
\autoref{tab:comparative-valley-splitting} compares the conduction band and the valence band valley splitting of Mo$_{0.75}$V$_{0.25}$Te$_2$ with several representative 2D systems. In the present work, Mo$_{0.75}$V$_{0.25}$Te$_2$ exhibited a maximum conduction band valley splitting of 54.4 meV and a maximum valence band valley splitting of 160.8 meV. These values demonstrated that V alloying, together with external strain engineering, generated a large and tunable valley polarization in MoTe$_2$.

Compared with proximity-induced heterostructures, the valley splitting of Mo$_{0.75}$V$_{0.25}$Te$_2$ was lower than that of MoTe$_2$/EuO, where conduction band and valence band valley splittings of 419 meV and 342 meV were reported, respectively~\cite{MoTe2}. Similarly, WS$_2$/h-VN showed larger conduction band and valence band valley splittings of 148 meV and 376 meV, respectively~\cite{ke2019large}. However, these systems relied on magnetic-proximity effects from magnetic substrates, 
which may increase structural complexity and influence the intrinsic electronic properties of the active TMD layer.

\begin{table*}[!t]
\caption{Comparative study of conduction band and valence band valley splitting in selected 2D systems}
\label{tab:comparative-valley-splitting}
\centering
\setlength{\tabcolsep}{6pt}
\renewcommand{\arraystretch}{1.18}
\small
\begin{tabularx}{\textwidth}{l Y Y l}
\toprule
\textbf{Structure} &
\textbf{$\Delta\textsubscript{K\textsubscript{+}K\textsubscript{-}}$ (meV) at conduction band} &
\textbf{$\Delta\textsubscript{K\textsubscript{+}K\textsubscript{-}}$ (meV) at valence band} &
\textbf{Reference} \\
\midrule

MoTe$_2$/EuO 
& 419 
& 342 
& ~~~\cite{MoTe2} \\

WS$_2$/h-VN 
& 148 
& 376 
& ~~~\cite{ke2019large} \\

WSe$_2$/CrI$_3$ 
& 0.26 
& 1.13 
& ~~~\cite{zhang2019valley} \\

2H-VSe$_2$ 
& 18 
& 78.2 
& ~~~\cite{abdollahi2023tuning} \\

SVSiN$_2$ 
& 4.32 
& 72.73 
& ~~~\cite{qi2024strain} \\

\textbf{Mo$_{0.75}$V$_{0.25}$Te$_2$} 
& \textbf{54.4} 
& \textbf{160.8} 
& \textbf{This work} \\

\bottomrule
\end{tabularx}
\end{table*}

On the other hand, Mo$_{0.75}$V$_{0.25}$Te$_2$ showed considerably stronger valley splitting than several previously reported 2D magnetic and ferrovalley systems. For example, WSe$_2$/CrI$_3$ exhibited relatively small conduction band and valence band valley splittings of 0.26 meV and 1.13 meV, respectively~\cite{zhang2019valley}. The valley splitting of Mo$_{0.75}$V$_{0.25}$Te$_2$ was also larger than that of 2H-VSe$_2$, where the conduction band and valence band valley splittings were 18 meV and 78.2 meV, respectively~\cite{abdollahi2023tuning}. Moreover, compared with SVSiN$_2$, which showed conduction band and valence band valley splittings of 4.32 meV and 72.73 meV, respectively~\cite{qi2024strain}, the present alloy exhibited a much stronger valley-polarized response at both band edges.

Although Mo$_{0.75}$V$_{0.25}$Te$_2$ did not exceed the giant valley splitting reported in some magnetic-substrate-based heterostructures, it outperformed several intrinsic and low-dimensional magnetic valleytronic systems. More importantly, the large valence band valley splitting of 160.8 meV and conduction band valley splitting of 54.4 meV were achieved in a V-alloyed MoTe$_2$ monolayer without constructing a separate magnetic heterostructure. This highlighted Mo$_{0.75}$V$_{0.25}$Te$_2$ as a promising and structurally simple platform for tunable valley devices.

\newcolumntype{Y}{>{\centering\arraybackslash}X}

\section{Conclusion}

In this work, we investigated the structural stability, electronic properties, and valley polarization of V-alloyed MoTe$_2$ monolayer, Mo$_{0.75}$V$_{0.25}$Te$_2$, using first-principles DFT calculations. Substitution of one Mo atom by a V atom in a ($2 \times 2 \times 1$) MoTe$_2$ supercell introduced magnetic exchange interaction into the host lattice, while the negative formation energy of $-0.84$ eV and absence of imaginary phonon modes confirmed the energetic and dynamical stability of the alloyed structure. Pristine MoTe$_2$ showed SOC-induced spin splitting but no valley polarization because the K+ and K-- valleys remained degenerate. In contrast, V alloying broke the valley degeneracy through the combined effect of magnetic exchange interaction and SOC, producing spontaneous valley polarization of 37.3 meV in the conduction band and 78.2 meV in the valence band. The valley polarization was further enhanced by external electric field and biaxial strain, reaching a maximum valence-band splitting of 132.8 meV under a transverse electric field of 0.4~\VperA\hspace{0.8mm} along the crystal $c$ axis and 160.8 meV under 4\% biaxial tensile strain, while the maximum conduction-band splitting reached 54.4 meV under 2\% compressive strain. These results demonstrate that Mo$_{0.75}$V$_{0.25}$Te$_2$ is a stable and highly tunable 2D valleytronic material, making V-alloyed MoTe$_2$ a promising platform for future valleytronic sensors, switches, and reconfigurable low-power electronic devices.






\printcredits
\section*{Declaration of competing interest}
The authors declare the absence of any financial or personal relationships that might have influenced the research presented in
this study.

\section*{Data availability statement}
{The data underlying the findings of this study are not publicly shared but available from the authors upon reasonable request.}

\section*{Acknowledgements}
M.M.I., V.C., M.N.A.D., and A.Z. gratefully acknowledge the Bangladesh University of Engineering and Technology (BUET) for providing computational resources and technical support. 

\appendix
\section{Supplementary material}
Supplementary material includes Figs. (S1–S6) showing additional information. It can be found online at https://....


\bibliography{cas-refs}
\balance

\vskip10pt
\clearpage
\appendix
\appendixpage

\section{Additional structural details}
\begin{figure*}[h]
\centering
\includegraphics[width=\textwidth]{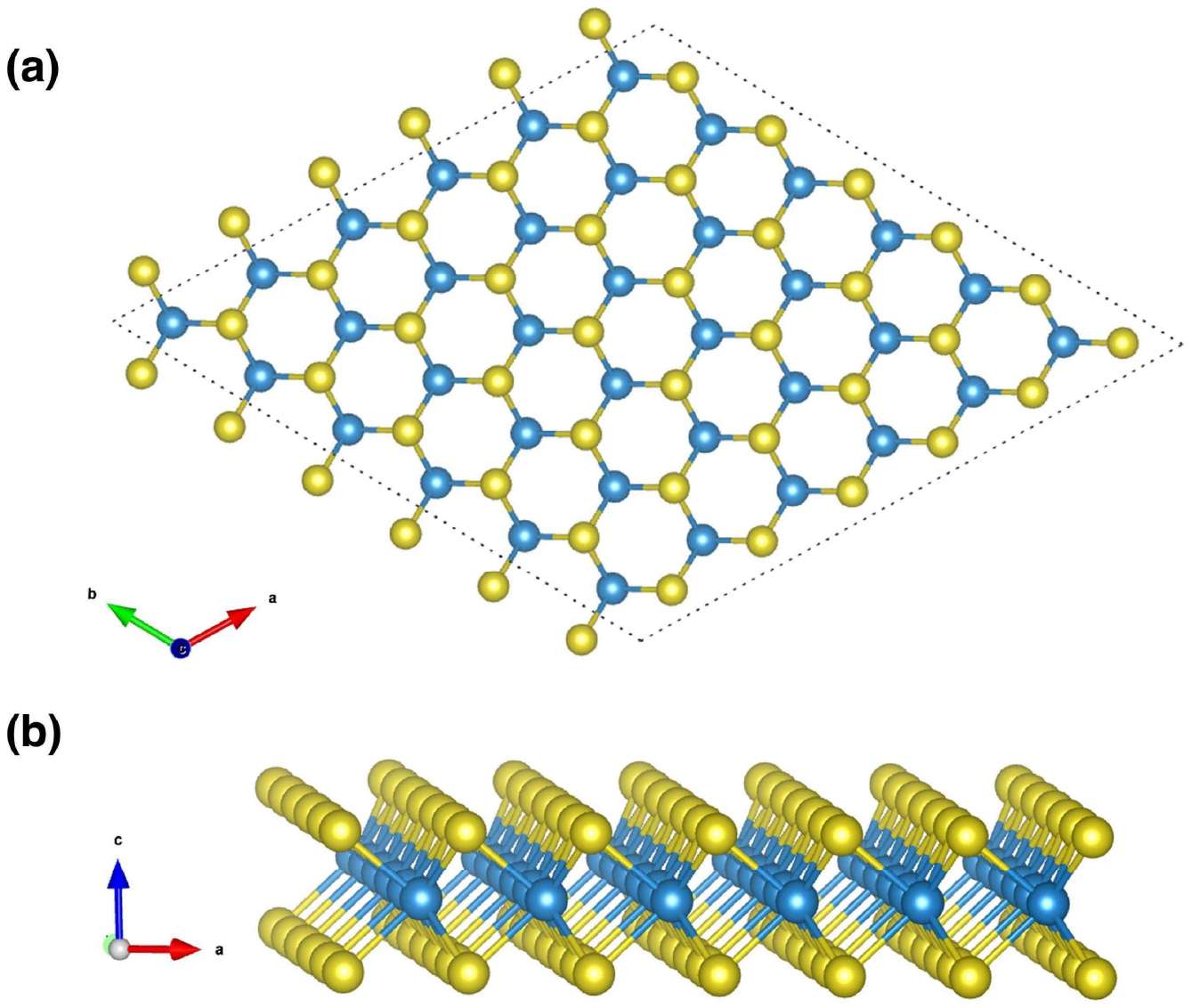}
\caption{Two dimensional pristine MoTe\textsubscript{2} with hexagonal symmetry. (a) Top view, (b) lateral view. Blue atoms are Mo, and yellow atoms are Te.
}
\label{Supp1}
\end{figure*}

\autoref{Supp1} shows the two dimensional (2D)  pristine MoTe\textsubscript{2} with hexagonal symmetry. \autoref{Supp1} (a) demonstrates the top view, and \autoref{Supp1} (b) demonstrates the side view. There are total 108 atoms. Blue atoms are Mo, and yellow atoms are Te. Among them, 72 atoms are Te, and 36 atoms are Mo. So, \% of Te = 66.67\%, and \% of Mo = 33.33\%. This ratio matches with the formula MoTe\textsubscript{2}.

\clearpage
\begin{figure*}[h]
\centering
\includegraphics[width=\textwidth]{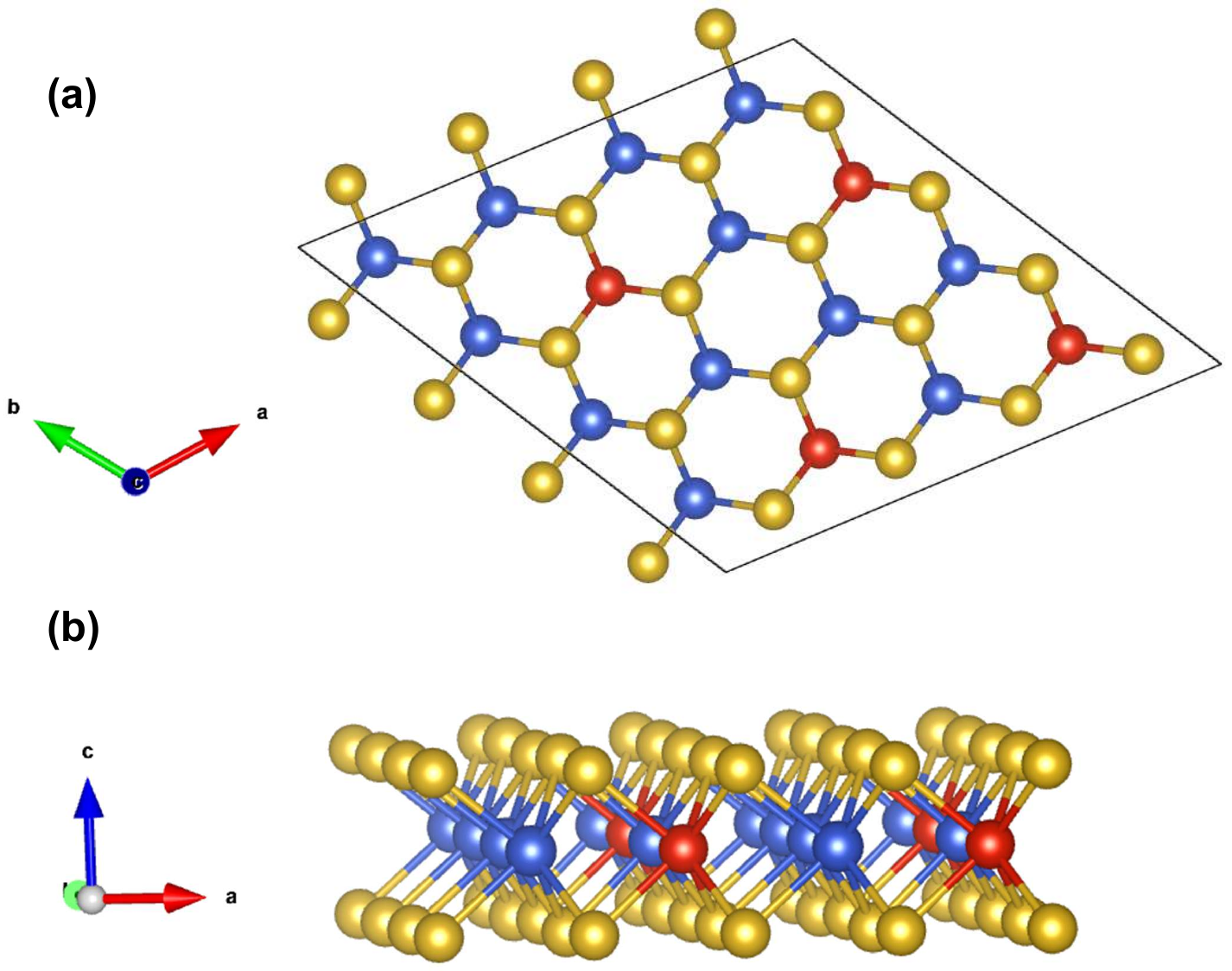}
\caption{Two dimensional V alloyed MoTe\textsubscript{2} (Mo\textsubscript{0.75}V\textsubscript{0.25}Te\textsubscript{2}) with hexagonal symmetry. (a) Top view, (b) lateral view. Blue atoms are Mo, red atoms are V, and yellow atoms are Te.}
\label{Supp2}
\end{figure*}

\autoref{Supp2} demonstrates the monolayer of V alloyed MoTe\textsubscript{2}      (Mo\textsubscript{0.75}V\textsubscript{0.25}Te\textsubscript{2}). \autoref{Supp2} (a) demonstrates the top view, and \autoref{Supp2} (b) demonstrates the side view. There are total 48 atoms. Among them, 32 atoms are Te. 12 atoms are Mo, and 4 atoms are V. So, \% of Te = 66.67\%, \% of Mo = 25\%, and \% of V = 08.33\%. This ratio matches with the formula Mo\textsubscript{0.75}V\textsubscript{0.25}Te\textsubscript{2} or Mo\textsubscript{3}VTe\textsubscript{8}.

\clearpage
\section{Stability of pristine MoTe\textsubscript{2}}
\begin{figure*}[h]
\centering
\includegraphics[width=0.8\textwidth]{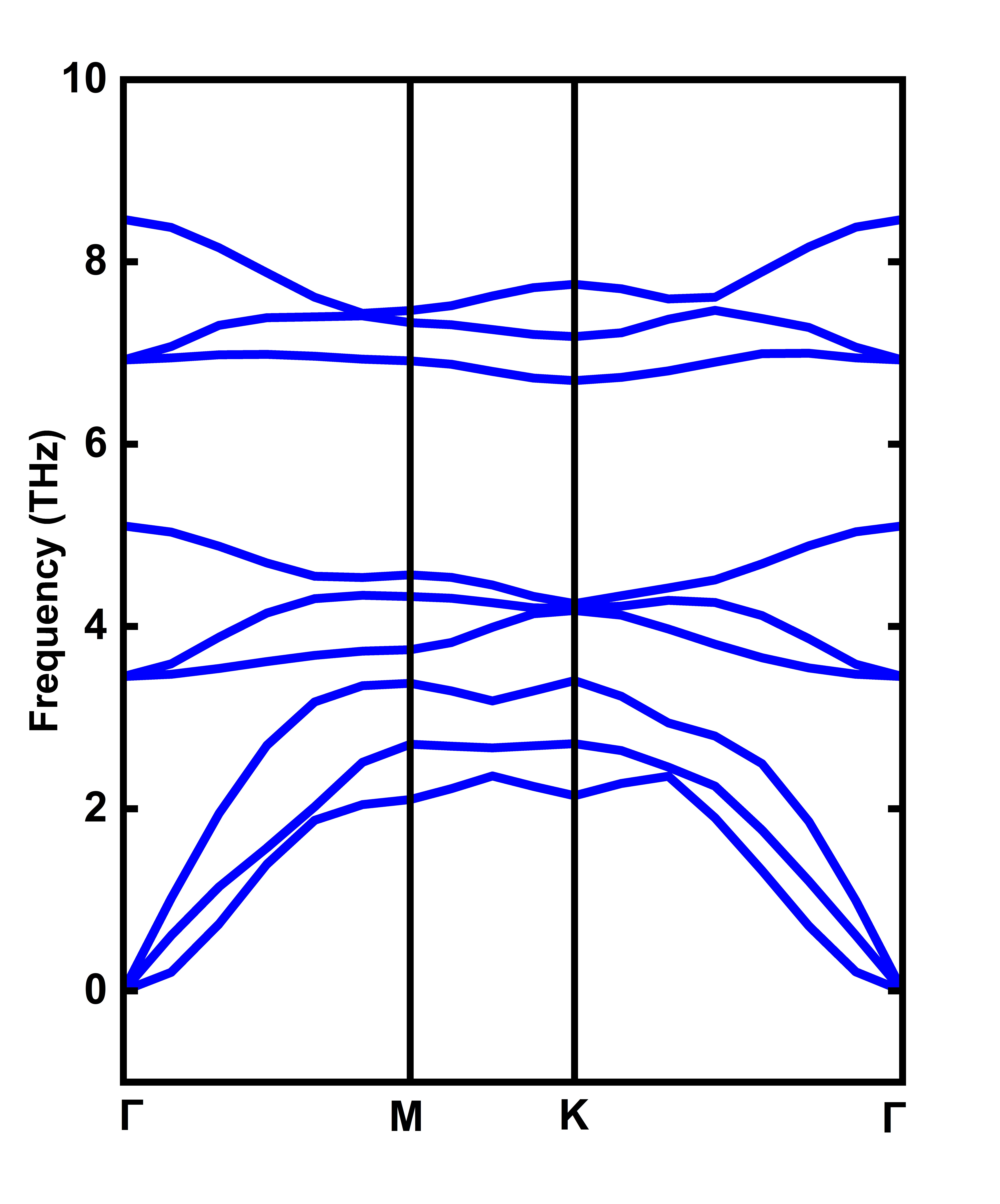}
\caption{Phonon band dispersion curve of pristine MoTe\textsubscript{2}.}
\label{Supp3}
\end{figure*}
For the investigation of dynamic stability, we conducted the phonon band
calculation of the Mo\textsubscript{2} structure. From the phonon dispersion curve shown in \autoref{Supp3}, we found that there is no imaginary phonon frequency for this structure. Therefore, the absence of imaginary phonon frequencies confirms the dynamic stability of MoTe\textsubscript{2}.


\clearpage
\section{Stability of pristine VTe\textsubscript{2}}
\begin{figure*}[h]
\centering
\includegraphics[width=0.8\textwidth]{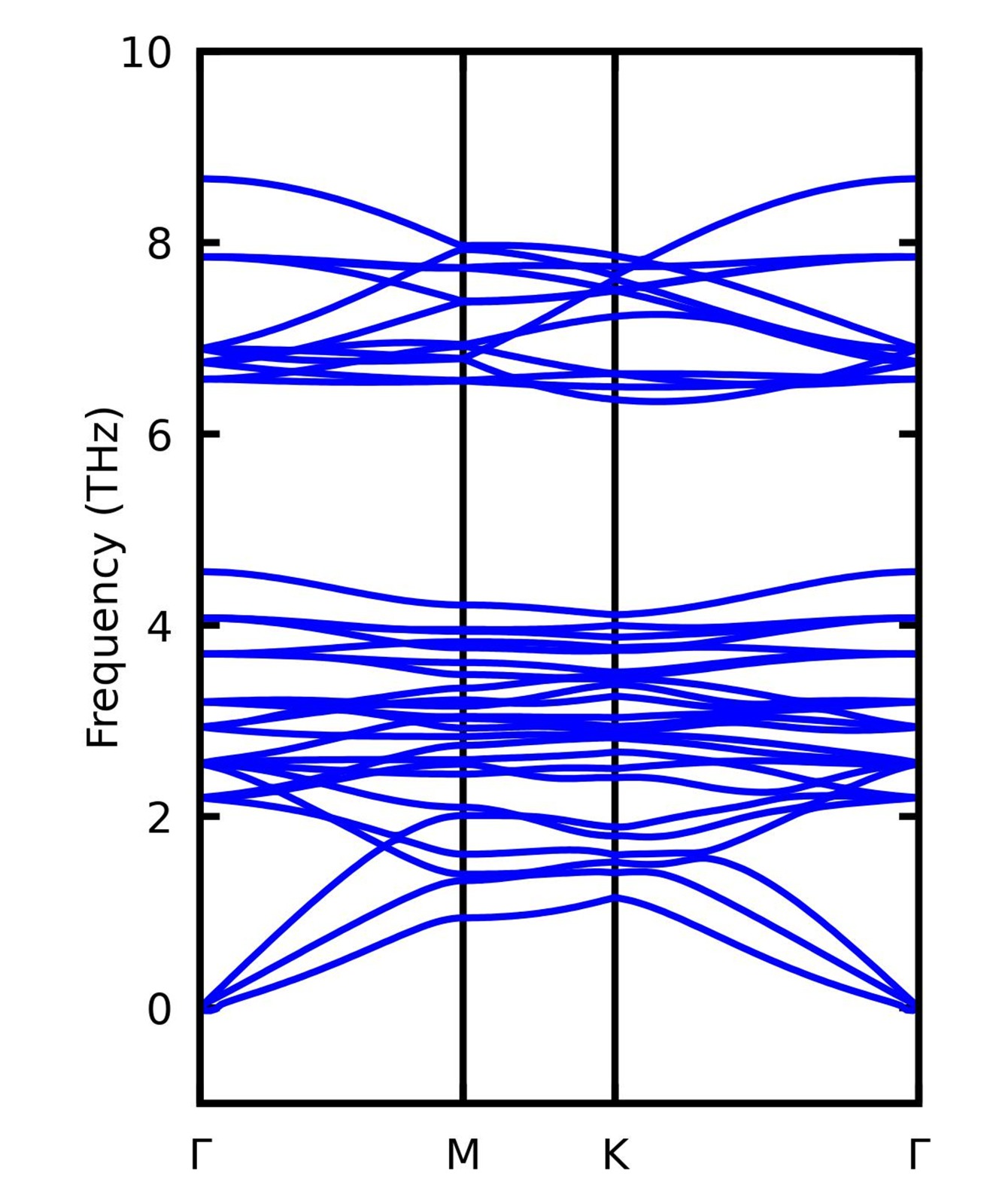}
\caption{Phonon band dispersion curve of pristine VTe\textsubscript{2}.}
\label{Supp4}
\end{figure*}

For the investigation of dynamic stability, we conducted the phonon band
calculation of the VTe\textsubscript{2} structure. From the phonon dispersion curve shown in \autoref{Supp4}, we found that there is no imaginary phonon frequency for this structure. Therefore, the absence of imaginary phonon frequencies confirms the dynamic stability of VTe\textsubscript{2}.

\clearpage
\section{Band structure of pristine VTe\textsubscript{2} without SOC}
\begin{figure*}[h]
\centering
\includegraphics[scale=0.48]
{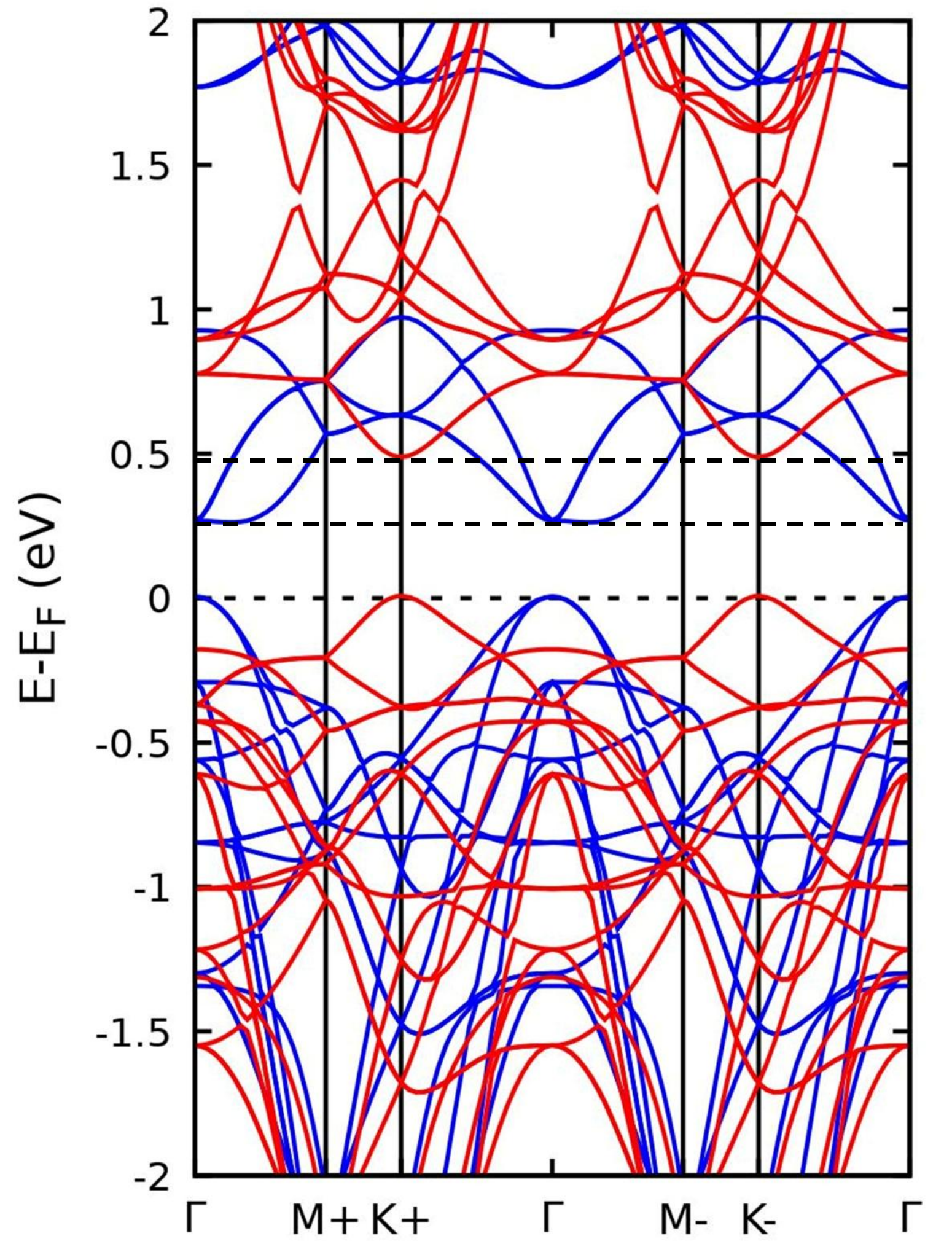}
\caption{Spin polarized band structure of pristine VTe\textsubscript{2} without SOC. Red color indicates spin-up and blue color
indicates spin-down band.}
\label{Supp5}
\end{figure*}

We studied the electronic properties of the pristine VTe\textsubscript{2} to match
with previous literature. We studied the spin-polarized band structure of pristine VTe\textsubscript{2} shown in \autoref{Supp5}. We explored that there is no spin splitting and no valley splitting at both the conduction bands and the valence
bands; the up-spin and the down-spin bands are degenerate. The band gap of pristine VTe\textsubscript{2} was found to be 0.25 eV. We also noticed that the Fermi level is at the edge of the valence band maxima (VBM).

\clearpage
\section{Band structure of pristine VTe\textsubscript{2} with SOC}
\begin{figure*}[h]
\centering
\includegraphics[scale=0.48]
{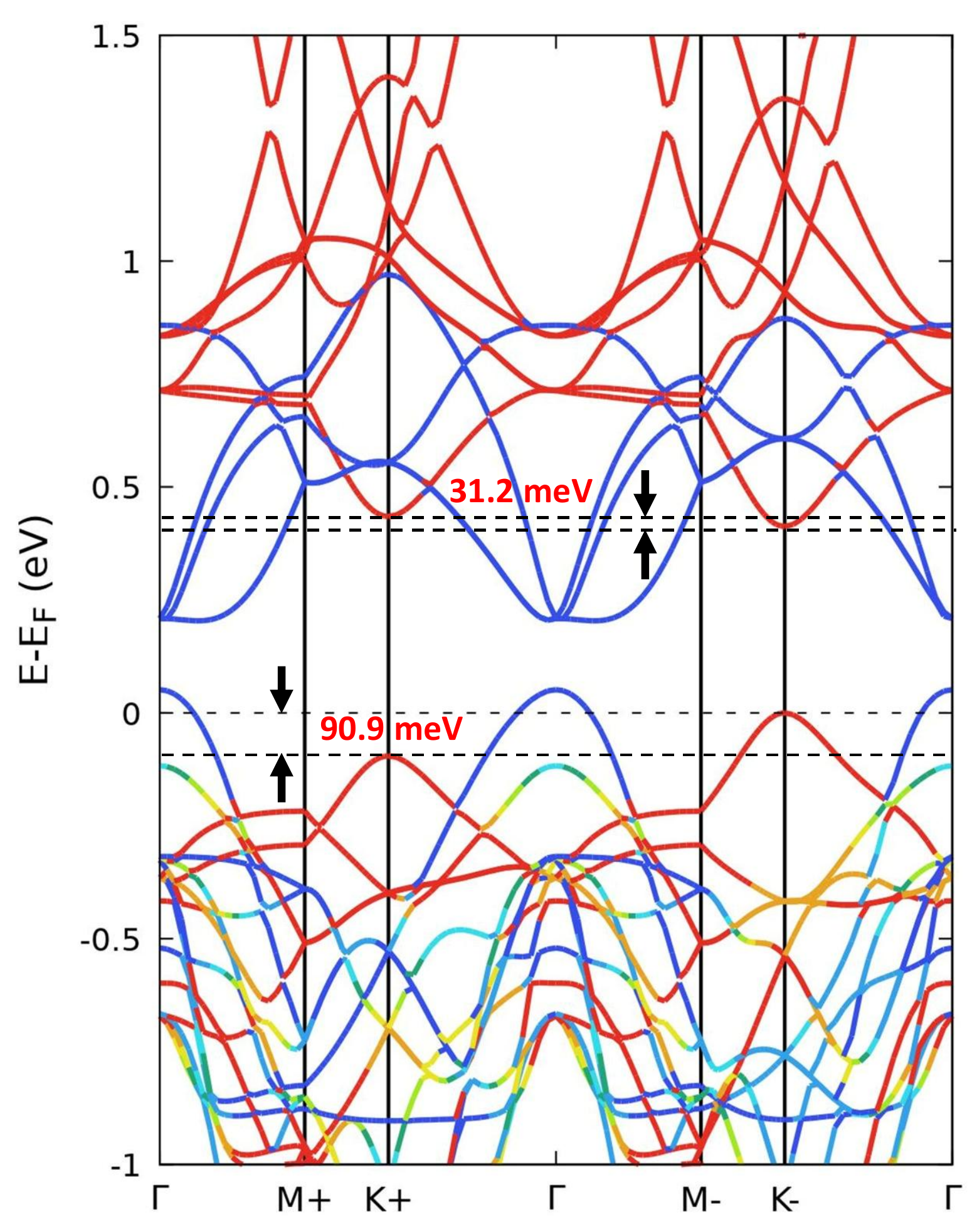}
\caption{Spin polarized band structure of pristine VTe\textsubscript{2} with SOC. Red color indicates spin-up and blue color
indicates spin-down band.}
\label{Supp6}
\end{figure*}

Then we studied the band structure of pristine VTe\textsubscript{2} with SOC effect. When SOC was introduced, some interesting phenomena appeared.
At first the band gap decreased to 0.18 eV after including SOC. Then, we saw valley
splitting in both the conduction band and the valence band at the K+ and the K-- points, although the valence band had the larger amount of valley splitting. In the conduction band, we determined a valley splitting of 31.2 meV, whereas a valley splitting of 90.9 meV was seen in the valence band.

\end{document}